\documentclass[pre,aps,twocolumn,superscriptaddress]{revtex4-2}

\usepackage{amssymb,amsfonts,amsmath}

\usepackage{verbatim} 
\usepackage{array} 
\usepackage{multirow} 
\usepackage{enumerate} 

\usepackage{graphicx}
\usepackage{tikz}
\usetikzlibrary{shapes}
\usetikzlibrary{patterns}
\usetikzlibrary{angles, quotes}
\definecolor{bostonuniversityred}{rgb}{0.8, 0.0, 0.0}
\definecolor{chromeyellow}{rgb}{1.0, 0.65, 0.0}

\usepackage{appendix}

\usepackage{color}
\usepackage[normalem]{ulem}
\usepackage{slashed}

\newcommand{\rhotwo}{\rho^{(2)}}

\newcommand{\unit}{\vec{e}}
\newcommand{\unittensor}{\mathbb{I}} 
\newcommand{\joe}{\color{black}}

\renewcommand{\vec}[1]{\mathrm{\mathbf{#1}}}

\newcommand{\M}{{\bf M}}
\newcommand{\E}{{\bf E}}
\newcommand{\Om}{{\boldsymbol{\Omega}}}
\newcommand{\kap}{{\boldsymbol{\kappa}}}

\usepackage[final]{showlabels} 

\begin{document}


\title{
Superadiabatic dynamical density functional theory for\\ 
colloidal suspensions under homogeneous steady-shear
}

\author{S.~M. Tschopp}
\affiliation{Department of Physics, University of Fribourg, CH-1700 Fribourg, Switzerland}
\author{J.~M. Brader}
\affiliation{Department of Physics, University of Fribourg, CH-1700 Fribourg, Switzerland}
\date{\today}

\begin{abstract}
The superadiabatic dynamical density functional theory 
(superadiabatic-DDFT) 
is a promising new method for the study of colloidal 
systems out-of-equilibrium. 
Within this approach the viscous forces arising from interparticle interactions are accounted for in a natural way by treating explicitly the dynamics of the two-body correlations.
For bulk systems subject to spatially homogeneous shear 
we use the superadiabatic-DDFT framework to calculate the steady-state pair 
distribution function and the corresponding viscosity for 
low values of the shear-rate. 
We then consider a variant of the central approximation underlying 
this superadiabatic theory
and obtain an inhomogeneous generalization of a rheological bulk theory due to Russel and Gast. This paper thus establishes for the first time a connection between 
DDFT approaches, formulated to treat inhomogeneous 
systems, and existing work addressing nonequilibrium 
microstructure and rheology in bulk colloidal suspensions.

\end{abstract}

\maketitle

\section{Introduction}

Colloidal suspensions exhibit a rich variety of 
rheological behaviour, arising from the interplay between 
Brownian motion, solvent hydrodynamics and potential 
interactions \cite{CatesSoftBook,WagnerBook,
BraderReview}.
For example, the phenomena of shear thinning, 
shear thickening and yielding are relevant for many 
commercial products and industrial processes. 
In order to control and tune the rheological properties of a suspension 
for any specific application it is necessary to have 
an understanding of how the microscopic interactions 
between the constituents influence the macroscopic 
response \cite{Larson}.
The challenge for nonequilibrium statistical mechanics 
is to formulate robust and accurate first-principles theories 
based on tractable approximation schemes which capture the 
essential physics while remaining sufficiently simple 
for concrete calculations to be performed.  

For the most commonly studied situation, namely bulk suspensions under homogeneous shear, there currently exist 
a variety of microscopic approaches. 
Each of these aims to capture a particular aspect 
of the cooperative particle motion within some limited range 
of shear-rates and thermodynamic parameters, but fail to provide a unified global picture. 
While exact results can be obtained for low density systems at 
low shear-rate \cite{BatchelorBook,Bergenholtz,BradyMorris}, 
systems at intermediate \cite{Ohtsuki,Ronis,RusselGast,WagnerRussel,BradyScaling,
BradyMorris,LionbergerRusselNoHydro,LionbergerRusselHydro,
Szamel,LionbergerRusselReview,MorrisReview,Russel_1989,banetta} or high densities \cite{FuchsCates2002,BraderRheo1,BraderRheo2,FuchsCates2009,
BraderRheo2012} invoke a diverse range of approximate closure relations to account for the correlated motion of the particles. 
Inhomogeneous systems, for which the density and shear-rate vary in space, are more challenging to treat theoretically and appropriate closure relations, which correctly capture the coupling between gradients in shear-rate and in density, 
remain under development \cite{MorrisMigration,BraderKruger,BraderKrugerSedimentation,AerovKrueger,BraderScacchi,FlowConcentration}.

The clearest path to a first-principles theory of suspension 
rheology is to focus on the particle correlation functions   
and then integrate these to obtain the macroscopic 
rheological properties of interest. 
This makes it possible to connect the macroscopic constitutive relations to the nonequilibrium microstructure 
of the system and thus gain microscopic insight into the 
physical mechanisms at work \cite{BatchelorBook,WagnerBook}. 
For systems with pairwise additive interparticle interactions the two-body correlations are the primary objects of interest. 
In the absence of hydrodynamic interactions, these enable full calculation of the stress tensor which is the key quantity of interest in rheology.
Additional motivation to develop theories which ‘look inside’
the flowing system is provided by developments in the visualization and tracking of particle motion in experiments (confocal microscopy) \cite{MorrisMigration,Crocker96,Prasad2007,Besseling2009}, together
with the detailed information provided by computer simulations  of model systems under flow \cite{FossBrady2000,Sierou2001,Banchio2003,
BradyTripletSimulations}.

Theoretical studies of inhomogeneous fluids, both in- and out-of-equilibrium, are primarily based on the spatially varying one-body density alone and do not usually involve directly the 
inhomogeneous two-body correlations (although exceptions do 
exist \cite{tschopp1,AttardSpherical,Kjellander2012,tschopp4,tschopp5}).  
This is a major difference between the standard dynamical density functional theory (standard DDFT) of inhomogeneous fluids \cite{Marconi98,ArcherEvansDDFT}
and the aforementioned approaches to bulk colloidal rheology.
When applied to bulk systems subject to homogeneous shear flow the standard DDFT does not present, as is, a useful framework, since the one-body density is not affected by the shear and remains constant in time. There have nevertheless been attempts to supplement the one-body equation of standard DDFT with additional empirical correction terms to avert this issue \cite{BraderKruger,BraderKrugerSedimentation,Dzubiella2003}. 

However, a new DDFT framework has recently been developed, 
the so-called superadiabatic-DDFT \cite{tschopp4}, 
which treats explicitly the dynamics of the two-body correlations. 
This scheme presents a significant improvement in describing the dynamics of inhomogenous fluids in the presence of time-dependent external potentials by accounting, via the two-body correlations, 
for the structural rearrangement of the particles as the system flows \cite{tschopp4,tschopp5}.
The superadiabatic-DDFT (composed of a pair of coupled equations for the one- and two-body density) per construction avoids the shortcomings of standard DDFT, since its equation for the two-body density is affected by shear and can thus be directly used to study bulk homogeneous systems.
By predicting the shear-induced distortion of the pair correlations, superadiabatic-DDFT allows to obtain the viscosity as an output of the theory. 
This is not a quantity reachable in any standard DDFT treatment of the problem, but is a very relevant bridge to the world of rheology.




In this paper we apply superadiabatic-DDFT to 
bulk systems under steady-shear flow and investigate 
its predictions for the shear-distorted pair distribution 
function and the low-shear viscosity. 
In addition, 
we show that a variation of the superadiabatic-DDFT reproduces, 
in the bulk limit, an early theory due to Russel and Gast \cite{RusselGast} and thus provides 
a generalization of their approach to the case of inhomogeneous fluids in external fields.  
This establishes a clear connection between existing 
microscopic theories of bulk colloidal rheology and 
DDFT approaches to the dynamics of 
inhomogeneous fluids. 
The numerical predictions of this Russel-Gast-type 
approach are compared with those of supradiabatic-DDFT. 

\section{Superadiabatic-DDFT}\label{super-DDFT}

\subsection{General framework}\label{super-DDFT-general}


The superadiabatic-DDFT, presented in detail in reference \cite{tschopp4}, consists of a pair of differential 
equations for the coupled time-evolution of the one- and two-body densities. 
It is applicable to systems with pairwise interparticle interactions. 
The first equation of superadiabatic-DDFT is the exact expression for the one-body density
\begin{align} \label{one-body exact}
\frac{1}{D_0}  \frac{\partial \rho(\vec{r}_1,t)}{\partial t} =& \,\nabla_{\vec{r}_1} \!\!\cdot \!\Bigg(\!\nabla_{\vec{r}_1} \rho(\vec{r}_1,t) + \rho(\vec{r}_1,t) \nabla_{\vec{r}_1} \beta V_{\text{ext}}(\vec{r}_1,t) \notag\\
&+ \int d \vec{r}_2 \, \rho^{(2)}(\vec{r}_1,\vec{r}_2,t) \nabla_{\vec{r}_1} \beta \phi(r_{12}) \!\Bigg), 
\end{align}
where $\beta\!=\!(k_BT)^{-1}$, $D_0$ is the diffusion coefficient,
$\phi$ is the interparticle pair potential, 
$r_{12}\!=\!|\vec{r}_1\!-\!\vec{r}_2|$ and $V_{\text{ext}}$ is a time-dependent external potential. 
The second equation is an approximate equation of motion for the two-body density, given by
\begin{align} \label{two body adiabatic}
&\frac{1}{D_0} \, \frac{\partial \rho^{(2)}(\vec{r}_1,\vec{r}_2,t)}{\partial t} = \\
&\sum_{i=1,2} \nabla_{\vec{r}_i} 
\!\cdot\! \Bigg( \!\nabla_{\vec{r}_i} \rho^{(2)}_{\text{\,sup}}(\vec{r}_1,\vec{r}_2,t)
\!+\! \rho^{(2)}_{\text{\,sup}}(\vec{r}_1,\vec{r}_2,t) \nabla_{\vec{r}_i} \beta \phi(r_{12}) \notag\\
&\!\!\!+\! \rho^{(2)}(\vec{r}_1,\vec{r}_2,t) \nabla_{\vec{r}_i} \beta V_{\text{ext}}(\vec{r}_i)
\!-\! \rho^{(2)}_{\text{ad}}(\vec{r}_1,\vec{r}_2,t) \nabla_{\vec{r}_i} \beta V_{\text{ad}}(\vec{r}_i,t) \!\!\Bigg) , \notag
\end{align}
where the superadiabatic contribution to 
the two-body density is defined according to
\begin{equation} \label{rho sup}
\rho^{(2)}_{\, \text{sup}}(\vec{r}_1,\vec{r}_2,t) \!\equiv\! \rho^{(2)}(\vec{r}_1,\vec{r}_2,t)-\rho^{(2)}_{\text{ad}}(\vec{r}_1,\vec{r}_2,t).
\end{equation}
The adiabatic two-body density, $\rho^{(2)}_{\text{ad}}$,
is obtained by evaluating the equilibrium two-body density functional at the instantaneous one-body density
\begin{equation} \label{rho2 ad}
\rho^{(2)}_{\text{ad}}(\textbf{r}_1,\textbf{r}_2,t)
\equiv
\rho^{(2)}_{\text{eq}}(\textbf{r}_1,\textbf{r}_2;
[\rho(\textbf{r},t)]). 
\end{equation}
The adiabatic potential, $V_{\text{ad}}$, appearing in 
\eqref{two body adiabatic} generates a fictitious external force which stabilizes the adiabatic system.  
This is obtained from the Yvon-Born-Green (YBG) relation 
of equilibrium 
statistical mechanics \cite{Hansen06,mcquarrie},
\begin{multline} \label{V ad} 
-\nabla_{\vec{r}_1} V_{\text{ad}}(\vec{r}_1,t) 
\equiv
k_BT\,\nabla_{\vec{r}_1} \ln \rho(\vec{r}_1,t) 
\\ 
+ \int d \vec{r}_2 \, 
\frac{\rho_{\text{ad}}^{(2)}(\vec{r}_1,\vec{r}_2,t)}
{\rho(\vec{r}_1,t)}
\nabla_{\vec{r}_1}\phi(r_{12}), 
\end{multline}
applied to the nonequilibrium system.
We note that the approximate equation for the two-body density 
\eqref{two body adiabatic} becomes exact in the low density limit.

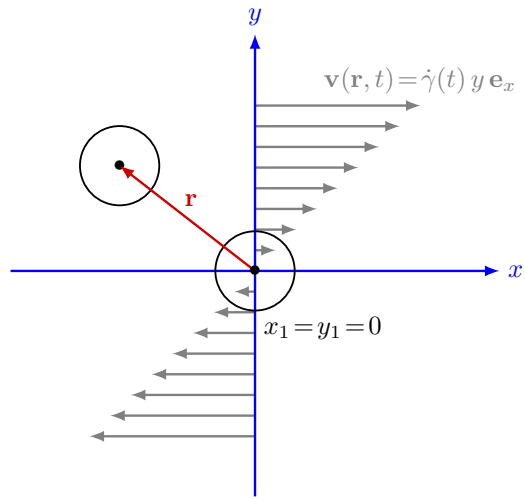
\begin{figure}[!t]
\hspace*{-0.5cm}
\begin{center}
\begin{tikzpicture}
\coordinate (origine) at (0,0);
\coordinate (particle 1) at (origine);
\coordinate (particle 2) at (-1.8,1.4);
\coordinate (y axes upper end) at (0,3.15);
\coordinate (y axes lower end) at (0,-3);
\coordinate (x axes left end) at (-3.25,0);
\coordinate (x axes right end) at (3.25,0);
\coordinate (label r12) at (-0.9+0.05,0.7+0.05);
\coordinate (label v) at (0.8,2.2+0.05);
\coordinate (shear line 1 left) at (0,2.2);
\coordinate (shear line 1 right) at (2.2,2.2);
\coordinate (shear line 2 left) at (0,1.925);
\coordinate (shear line 2 right) at (1.925,1.925);
\coordinate (shear line 3 left) at (0,1.65);
\coordinate (shear line 3 right) at (1.65,1.65);
\coordinate (shear line 4 left) at (0,1.375);
\coordinate (shear line 4 right) at (1.375,1.375);
\coordinate (shear line 5 left) at (0,1.1);
\coordinate (shear line 5 right) at (1.1,1.1);
\coordinate (shear line 6 left) at (0,0.825);
\coordinate (shear line 6 right) at (0.825,0.825);
\coordinate (shear line 7 left) at (0,0.55);
\coordinate (shear line 7 right) at (0.55,0.55);
\coordinate (shear line 8 left) at (0,0.275);
\coordinate (shear line 8 right) at (0.275,0.275);
\coordinate (shear line 16 left) at (0,-2.2);
\coordinate (shear line 16 right) at (-2.2,-2.2);
\coordinate (shear line 15 left) at (0,-1.925);
\coordinate (shear line 15 right) at (-1.925,-1.925);
\coordinate (shear line 14 left) at (0,-1.65);
\coordinate (shear line 14 right) at (-1.65,-1.65);
\coordinate (shear line 13 left) at (0,-1.375);
\coordinate (shear line 13 right) at (-1.375,-1.375);
\coordinate (shear line 12 left) at (0,-1.1);
\coordinate (shear line 12 right) at (-1.1,-1.1);
\coordinate (shear line 11 left) at (0,-0.825);
\coordinate (shear line 11 right) at (-0.825,-0.825);
\coordinate (shear line 10 left) at (0,-0.55);
\coordinate (shear line 10 right) at (-0.55,-0.55);
\coordinate (shear line 9 left) at (0,-0.275);
\coordinate (shear line 9 right) at (-0.275,-0.275);
\draw[->, >=latex, gray, line width=0.9] (shear line 1 left) -- (shear line 1 right);
\draw[->, >=latex, gray, line width=0.9] (shear line 2 left) -- (shear line 2 right);
\draw[->, >=latex, gray, line width=0.9] (shear line 3 left) -- (shear line 3 right);
\draw[->, >=latex, gray, line width=0.9] (shear line 4 left) -- (shear line 4 right);
\draw[->, >=latex, gray, line width=0.9] (shear line 5 left) -- (shear line 5 right);
\draw[->, >=latex, gray, line width=0.9] (shear line 6 left) -- (shear line 6 right);
\draw[->, >=latex, gray, line width=0.9] (shear line 7 left) -- (shear line 7 right);
\draw[->, >=latex, gray, line width=0.9] (shear line 8 left) -- (shear line 8 right);
\draw[->, >=latex, gray, line width=0.9] (shear line 9 left) -- (shear line 9 right);
\draw[->, >=latex, gray, line width=0.9] (shear line 10 left) -- (shear line 10 right);
\draw[->, >=latex, gray, line width=0.9] (shear line 11 left) -- (shear line 11 right);
\draw[->, >=latex, gray, line width=0.9] (shear line 12 left) -- (shear line 12 right);
\draw[->, >=latex, gray, line width=0.9] (shear line 13 left) -- (shear line 13 right);
\draw[->, >=latex, gray, line width=0.9] (shear line 14 left) -- (shear line 14 right);
\draw[->, >=latex, gray, line width=0.9] (shear line 15 left) -- (shear line 15 right);
\draw[->, >=latex, gray, line width=0.9] (shear line 16 left) -- (shear line 16 right);
\draw[gray] (label v) node[above right] {\normalsize $\vec{v}(\vec{r},t) \!=\! \dot{\gamma}(t) \, y \, \unit_x$};
\draw[->, >=latex, blue, line width=0.9] (y axes lower end) -- (y axes upper end);
\draw[->, >=latex, blue, line width=0.9] (x axes left end) -- (x axes right end);
\draw[blue] (y axes upper end) node[above] {\normalsize $y$};
\draw[blue] (x axes right end) node[right] {\normalsize $x$};
\draw[black] (0,-0.5) node[below right] {\normalsize $x_1 \!=\! y_1 \!=\! 0$};
\draw (particle 2) node {$\bullet$};
\draw[line width=0.7] (particle 1) circle (15pt);
\draw[line width=0.7] (particle 2) circle (15pt);
\draw[->, >=latex, bostonuniversityred, line width=0.9] (particle 1) -- (particle 2);
\draw (particle 1) node {$\bullet$};
\draw[bostonuniversityred] (label r12) node[above] {\normalsize $\vec{r}$};
\end{tikzpicture}
\end{center}
\caption{Sketch of the geometry. We choose particle 1 as the origin of our
cartesian coordinate system ($\vec{r}_1$ is thus implicitly fixed equal to zero).
}

\label{fig geometry coordinates}

\end{figure}

\subsection{Low shear-rate solutions for the pair distribution function in bulk}
\label{super-DDFT-solutions} 

We now specialize to a bulk system, for which the external potential is set equal to zero 
and the one-body density becomes constant, 
$\rho(\vec{r},t)\!\rightarrow\!\rho_b$. 
In this case the adiabatic two-body density is both isotropic 
and translationally invariant.  
Consequently, the integral term in equation \eqref{V ad} vanishes and the adiabatic potential becomes constant 
in space. 
This constant can then be set equal to zero without 
loss of generality.
To exploit the translational invariance of the system
we now define the following relative and absolute coordinates
\begin{equation}
\vec{r}=\vec{r}_1-\vec{r}_2\,,\quad
\vec{R}=\vec{r}_1+\vec{r}_2,
\end{equation}
that then allow us to make the replacements
\begin{align}
\nabla_{\vec{r}_1}&\rightarrow\quad\nabla_{\vec{r}}=\nabla,
\\
\nabla_{\vec{r}_2}&\rightarrow \,-\nabla_{\vec{r}}=-\nabla,
\end{align}
where we henceforth drop the subscripts on the nabla. 
Due to translational invariance $\rhotwo$ will not depend on 
the absolute position coordinate. 

Since we wish to investigate systems under homogeneous flow we 
introduce the affine velocity field, $\vec{v}$, which, for 
shear applied in the direction of the $x$-axis, with shear-gradient in the $y$-direction, is given by
\begin{equation} \label{shear definition}
\vec{v}(\vec{r},t) = \dot{\gamma}(t) \, y \, \unit_x,
\end{equation}
where $\dot{\gamma}$ is the shear-rate 
(as illustrated in Fig.~\ref{fig geometry coordinates}).
From equation \eqref{two body adiabatic} we thus obtain the following equation of motion for the pair distribution function
\begin{multline}
\frac{\partial g(\vec{r},t)}{\partial t} = 
-2 D_0 \nabla\cdot\Big(
-\nabla g_{\text{sup}}(\vec{r},t) - 
g_{\text{sup}}(\vec{r},t)\nabla\beta\phi(r)
\Big)
\\
-\nabla\cdot\Big(
g(\vec{r},t)\,\vec{v}(\vec{r},t)
\Big),
\label{pair distribution eom}
\end{multline}
in which we note the emergence of the pair diffusion constant, $2D_0$, and where the nonequilibrium pair distribution function is given by
\begin{equation}
g(\vec{r},t)=\frac{\rhotwo(\vec{r},t)}{\rho_b^2}.
\end{equation}
Its superadiabatic component, which encodes the flow-induced distortion of the microstructure, 
is defined as 
\begin{equation}
g_{\text{sup}}(\vec{r},t) = g(\vec{r},t) - g_{\text{eq}}(r),
\label{split}
\end{equation}
where the equilibrium radial distribution function, 
$g_{\text{eq}}$, can be obtained from the interaction pair potential using an appropriate equilibrium theory (we will use an integral equation closure). 

Finding a solution to equation \eqref{pair distribution eom} for arbitrary values of $\dot{\gamma}$ and $\rho_b$ is a difficult task. 
Even in the low density limit considerable effort must be expended to obtain numerical solutions, due to the emergence 
of a boundary-layer in $g$ as the shear-rate is increased \cite{Bergenholtz}. 
We will henceforth restrict our attention to the special case of steady-shear, 
$\dot{\gamma}(t)\!\rightarrow\!\dot{\gamma}$, for which 
the time-independent, steady-state pair distribution function, $g(\vec{r},t)\!\rightarrow\!g(\vec{r})$, can be obtained (almost) analytically for 
arbitrary values of $\rho_b$.
In the steady-state, equation \eqref{pair distribution eom} reduces to
\begin{multline}
2 D_0 \nabla\cdot\Big(
\nabla g_{\text{sup}}(\vec{r}) + 
g_{\text{sup}}(\vec{r})\nabla\beta\phi(r)
\Big)
\\
-\nabla\cdot\Big(
g(\vec{r})\,\vec{v}(\vec{r})
\Big)=0.
\label{steady state}
\end{multline}
In equilibrium, $\vec{v}(\vec{r})\!=\!0$ and the pair-current, $\nabla g_{\text{sup}}(\vec{r}) + 
g_{\text{sup}}(\vec{r})\nabla\beta\phi(r)$, vanishes. 
This condition leads trivially to 
the solution $g_{\text{sup}}(\vec{r})\!=\!0$, which implies that 
$g(\vec{r})\!=\!g_{\text{eq}}(r)$, as expected.

To obtain a low shear-rate solution of the steady-state 
equation \eqref{steady state} 
we assume 
$g_{\text{sup}}$ to be a linear function of $\dot{\gamma}$. 
{\joe (This will be sufficient for our present purposes. However, we 
note that care should be exercised when assuming linearity, since boundary-layer formation 
can significantly complicate the picture \cite{Dhont_1989,banetta,riva,zaccone_intermediate}.)}
Substitution of \eqref{split} into \eqref{steady state} and 
neglecting terms quadratic and higher in $\dot{\gamma}$ 
then yields 
the following linearized steady-state condition
\begin{multline}
2 D_0 \nabla\cdot\Big(
\nabla g_{\text{sup}}(\vec{r}) + 
g_{\text{sup}}(\vec{r})\nabla\beta\phi(r)
\Big)
\\
-\nabla\cdot\Big(
g_{\text{eq}}(r)\,\vec{v}(\vec{r})
\Big)=0.
\label{steady state linear}
\end{multline}
In order to capture correctly the anisotropy induced by 
the flow we make the following ansatz \cite{Russel_1989}
\begin{equation}
g_{\text{sup}}(\vec{r})=
-\frac{1}{2 D_0}\left(
\frac{\vec{r}\cdot\bold{E}\cdot\vec{r}}{r^2}
\right)
e^{-\beta\phi(r)}f(r),
\label{ansatz}
\end{equation}
where we have introduced the rate-of-strain tensor, $\bold{E}$,  and the isotropic radial function $f(r)$.
The rate-of-strain tensor contains all relevant information about the affine flow field,
while the function $f(r)$ depends only on the interaction potential, the bulk density and the system dimensionality. 
(Additional information regarding the rate-of-strain tensor is 
provided in appendix \ref{rheology appendix}.)

Working through the substitution of expression \eqref{ansatz} into equation
\eqref{steady state linear}, as described in appendix \ref{radial balance equ appendix}, finally yields the radial balance equations, required to determine $f(r)$. 
These are given by
\begin{align}\label{radial balance eq}
\frac{dg_{\text{eq}}(r)}{dr} & \, \overset{\text{2D}}{=} \,
-\frac{1}{r^2} \frac{d}{dr} \Big( r \frac{d f(r)}{dr} e^{-\beta \phi(r)} \Big) \;\,+ \frac{4}{r^3} f(r) e^{-\beta \phi(r)}, \notag\\
\frac{dg_{\text{eq}}(r)}{dr} & \, \overset{\text{3D}}{=} \,
-\frac{1}{r^3} \frac{d}{dr} \Big( r^2 \frac{d f(r)}{dr} e^{-\beta \phi(r)} \Big) + \frac{6}{r^3} f(r) e^{-\beta \phi(r)},
\end{align}
in two- and three-dimensions, respectively.
The boundary conditions required to solve equations \eqref{radial balance eq} depend on the pair potential under consideration and are discussed in detail in appendix 
\ref{boundary conditions}.

Although we have chosen to focus on shear, we note that the expressions 
presented in this subsection remain valid for any incompressible, translationally invariant flow field and so could be applied directly to, e.g.~extensional flows.


\subsection{Low-shear viscosity}\label{low shear visc section}

For a bulk system in a steady-state, the nonequilibrium pair 
distribution function is related to the stress tensor 
according to the following exact expression 
\cite{BraderReview}
\begin{equation}\label{stress eq}
\boldsymbol{\sigma}=-k_BT\rho_b \unittensor 
+ \frac{1}{2}\rho_b^2\int d{\bf r}\,
\frac{\vec{r}\vec{r}}{r}
\left(\frac{d\phi(r)}{dr}\right)
g(\vec{r}),
\end{equation}
where $\unittensor$ is the unit tensor. 
Due to the isotropy of the equilibrium pair distribution function only $g_{\text{sup}}$ contributes to the off-diagonal stress tensor elements.

The interaction part of the viscosity, $\eta$, is 
obtained by dividing $\sigma_{\text{xy}}$ by the shear rate.  
In the limit of low shear-rate, 
$\dot{\gamma}\!\rightarrow\! 0 $,  equation \eqref{stress eq} gives the low shear 
viscosity, $\eta_0$.
Substituting equation \eqref{ansatz} into equation 
\eqref{stress eq} and using the appropriate form for the rate-of-strain tensor in shear flow, given by equation \eqref{E matrix}, then yields
\begin{equation}\label{viscosity eq}
\eta_0 = \frac{k_BT}{4 D_0}\,\rho_b^2\int d\vec{r}\,
\frac{x^2y^2}{r^3}
\left(\frac{d}{dr}e^{-\beta\phi(r)}\right)
f(r).
\end{equation}
Both the function $f$ and the integral in 
\eqref{viscosity eq} depend on the dimensionality of the system. 
We then find 
\begin{align}\label{explicit viscosity eq}
\eta_0 & \, \overset{\text{2D}}{=} \;
\frac{k_BT}{\pi D_0}\,\Phi_{\text{2D}}^2 
\int_0^{\infty} dr\,r^2
\left(\frac{d}{dr}e^{-\beta\phi(r)}\right)f(r), \notag\\
\eta_0 & \, \overset{\text{3D}}{=} \;
\frac{k_BT}{\pi D_0}\,\frac{12}{5}\,
\Phi_{\text{3D}}^2 
\int_0^{\infty} dr\,r^3
\left(\frac{d}{dr}e^{-\beta\phi(r)}\right)f(r),
\end{align}
in two- and three-dimensions, respectively. 
We have introduced the two-dimensional area fraction, 
$\Phi_{\text{2D}}\!=\!\pi\rho_b/4$, and three-dimensional 
volume fraction, $\Phi_{\text{3D}}\!=\!\pi\rho_b/6$, 
where all lengths have been non-dimensionalized using the 
characteristic diameter of the particles.  
If the function $f$ were known exactly, then equation 
\eqref{explicit viscosity eq} would yield the exact 
low-shear viscosity.  
Within the present approach the adiabatic approximation 
employed to close the two-body equation of motion \eqref{two body adiabatic} yields approximate radial balance 
equations \eqref{radial balance eq} for $f$, and thus an approximate low-shear viscosity.


For the well-studied case of low density hard-spheres in three-dimensions, equation \eqref{radial balance eq} has the 
analytic solution, $f(r)\!=\!1/(3r^3)$, and equation \eqref{explicit viscosity eq} recovers the quadratic 
term in the well-known low density expansion (see page 114 in reference \cite{WagnerBook}),
\begin{equation}\label{einstein eq}
\eta_0 = \eta_s \left(1 + \frac{5}{2}\Phi_{\text{3D}} 
+\frac{12}{5}\Phi_{\text{3D}}^2\right),
\end{equation}
which applies in the absence of hydrodynamic interactions between the particles.  
The first term in \eqref{einstein eq} is the solvent 
viscosity, $\eta_s$. 
The second term arises from the 
drag of the solvent on the surface of each individual 
sphere \cite{BradyEinstein}. 
{\joe The third term, which is exact for a system without 
hydrodynamic interactions, 
represents the influence of direct 
potential interactions between the particles and comes from evaluating the integral in equation \eqref{explicit viscosity eq} and then 
employing the Stokes-Einstein relation $k_BT/D_0\!=\!3\pi\eta_s$.}

\subsection{Numerical results}\label{super-DDFT-results}

To investigate the predictions of the superadiabatic-DDFT approach we will focus on the special 
case of hard-disks in two-dimensions,
which presents a phenomenology qualitatively similar to that of hard-spheres in three-dimensions, while remaining convenient for visualization of the distorted pair correlations in the $xy$-plane. 
Solution of the radial balance equation \eqref{radial balance eq} requires as input the hard-disk equilibrium radial distribution function, $g_{\text{eq}}$.
We obtain this quantity by employing the famous Percus-Yevick closure of the homogeneous Ornstein-Zernike 
equation \cite{Lado,Lado_disks}, known to be accurate for the 
hard-disk system at low and intermediate bulk densities.

\begin{figure}
\includegraphics[width=\linewidth]{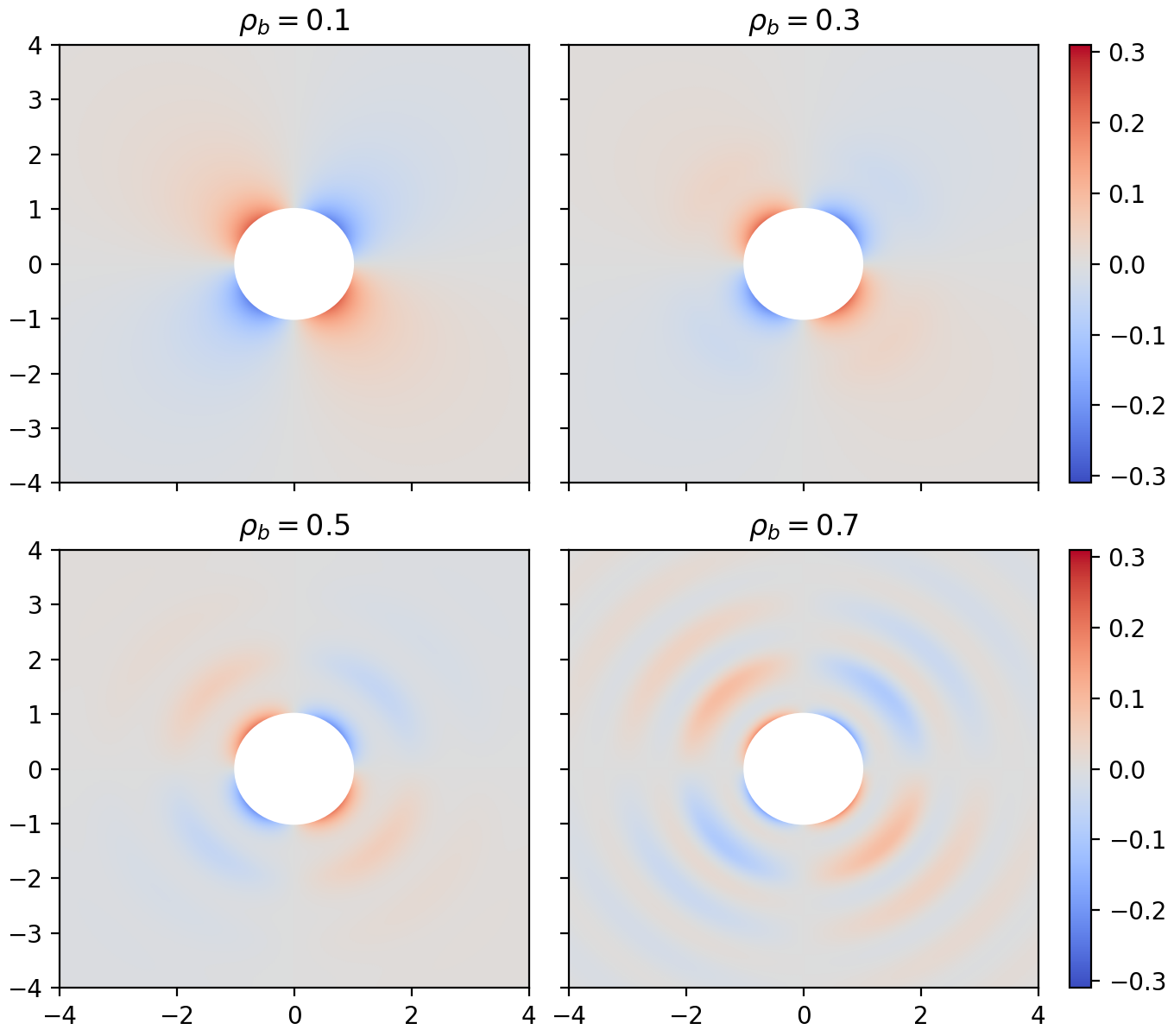}
\caption{\textbf{Superadiabatic-DDFT \normalsize{$g_{\text{sup}}$}}. 
For hard-disks in two-dimensions we show 
the superadiabatic contribution to the pair 
correlation function, $g_{\text{sup}}$, in units 
of $\dot{\gamma} d^2/2 D_0$. 
Since we consider only low shear-rates 
$g_{\text{sup}}$ exhibits quadipolar symmetry, 
see equation \eqref{ansatz}.   
Increasing the bulk density leads to  packing oscillations.
} 
\label{fig Sup-DDFT}
\end{figure}

In Fig.~\ref{fig Sup-DDFT}, we show scatter plots of the rescaled superadiabatic contribution to the distorted pair correlation function, namely
\begin{equation}
\frac{g_{\text{sup}}(x,y)}{\dot{\gamma} d^2/2 D_0} = -\frac{xy}{r^2}f(r),
\end{equation}
where $r\!=\!\sqrt{x^2+y^2}$ and $f(r)$ is obtained from the numerical integration of equation \eqref{radial balance eq}.
Results are shown for four different values of the bulk density, $\rho_{b}$, one per panel, and are valid at low shear-rates.
At the lowest considered bulk density, $\rho_{b}\!=\!0.1$, the radial function $f(r)$ obtained numerically is very close to its low-density limit value, $f(r)\!=\!1/2 r^2$, and no 
packing oscillations are visible in the resulting scaled $g_{\text{sup}}$.
In this panel we observe an accumulation of particles at contact, $r\!=\!d$, in the `compressional' quadrants, defined as the region where $\text{sign}(y)\!=\!-\text{sign}(x)$, and an opposite depletion effect in the `extensional' quadrants, where $\text{sign}(y)\!=\!\text{sign}(x)$. 
As the value of the bulk density is increased packing oscillations develop at larger values of $r$. These oscillations 
in $g_{\text{sup}}$ take both positive and negative values 
and are a nontrivial prediction of the radial balance equation 
\eqref{radial balance eq}.  
The packing structure of the full pair correlation function, 
$g(\vec{r})\!=\!g_{\text{eq}}(r)+g_{\text{sup}}(\vec{r})$, will thus differ from that observed in bulk as 
a consequence of the applied shear.
However, our numerical solutions reveal that the contact value of the radial function remains unaffected by changes in $\rho_{b}$ and is given by
\begin{equation} \label{contact value f}
f(r\!=\!d) = 1/(2d^2), \quad \forall \rho_b,
\end{equation}
in each of the panels shown in Fig.~\ref{fig Sup-DDFT}.

\begin{figure}
\includegraphics[width=\linewidth]{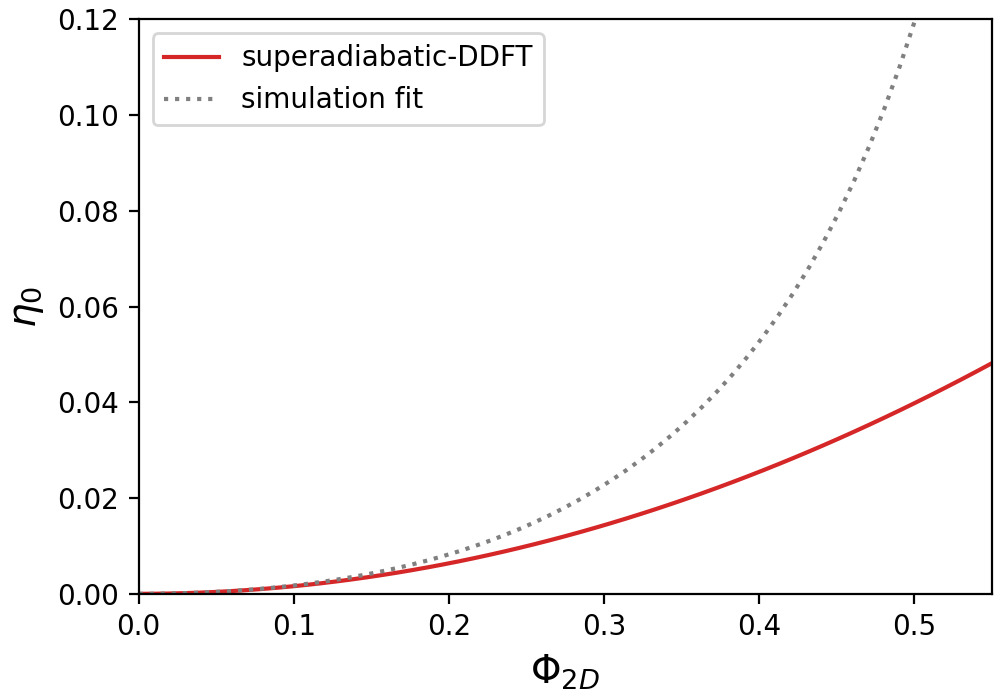}
\caption{\textbf{Low-shear viscosity}.
A comparison of $\eta_0$ from superadiabatic-DDFT, see equation \eqref{superDDFT viscosity}, with a fit to 
simulation data, taken from reference \cite{Reinhardt_2013}.
The low-shear viscosity from simulation diverges as the system 
approaches random close packing, whereas the 
superadiabatic-DDFT retains 
the low density limiting form for all values of the area fraction.
}
\label{fig Sup-DDFT viscosity}
\end{figure}

Knowledge of the nonequilibrium pair correlation function also enables calculation of
the interaction contribution to the low-shear viscosity, $\eta_0$, via equation \eqref{viscosity eq}.
Results obtained by solving the radial balance equation 
\eqref{radial balance eq} and evaluating the integral \eqref{viscosity eq}
for each chosen value of the bulk density
are shown in Fig.~\ref{fig Sup-DDFT viscosity} and compared with a fit to Brownian dynamics 
simulation data taken from reference \cite{Reinhardt_2013}. While both curves overlap for low values of 
$\rho_b$, the theoretical superadiabatic-DDFT curve strongly underestimates
$\eta_0$ relative to the simulation at higher densities.
We note that the simulation curve diverges as the system approaches random close packing, located at $\Phi_{\text{2D}}\!\approx\!0.82$ for monodisperse hard-disks.

Due to the density independence of the contact value \eqref{contact value f},
the above described route to obtaining the interaction contribution to the low-shear viscosity
recovers its low-density form, namely
\begin{equation}\label{superDDFT viscosity}
\eta_0 = \frac{k_BT d^2}{D_0}\frac{\Phi_{\text{2D}}^2}{2\pi}, \quad \forall \rho_b.
\end{equation}
Accounting for the influence of shear-flow on the bulk three-body density
would give corrections to higher-order in area fraction in equation \eqref{superDDFT viscosity}. 
However this mechanism is neglected per construction within the current approximation.  
{\joe The fact that we recover only the 
leading-order contribution \eqref{superDDFT viscosity} 
to the low-shear viscosity is thus not surprising and 
is consistent with application of the adiabatic approximation 
at the two-body level. 
It seems to us very likely that improved predictions for $\eta_0$ 
could be obtained from superadiabatic-DDFT by employing 
the `test-particle method', as we will discuss 
in section \ref{discussion}. 
}

\section{Russel-Gast-type theory}\label{section RG}

In the preceding section we analysed the properties of 
superadiabatic-DDFT for bulk systems subject to homogeneous 
shear. All predictions of the theory for the pair 
distribution function and viscosity arise from the adiabatic 
closure employed to arrive at equation \eqref{two body adiabatic}. 
In this section we show that there exists an alternative way to implement the adiabatic closure and that this results in a 
new approximate equation of motion for the inhomogeneous 
two-body density, different from equation \eqref{two body adiabatic}.   
In the bulk limit the new approximation reduces to an 
equation of motion for the pair distribution function 
first introduced in 1986 by Russel and Gast 
\cite{RusselGast} and which constituted one of the earliest 
theoretical approaches to the microstructure and rheology 
of colloidal suspensions at finite volume fraction. 
It thus seems appropriate to call our adiabatic closure 
of the equation of motion for the inhomogeneous two-body 
density a `Russel-Gast-type' approximation.



\subsection{Alternative adiabatic closure}\label{RG-closure}


Integrating the many-particle Smoluchowski equation over 
$N\!-\!2$ particle coordinates yields the following, 
formally exact equation of motion for the two-body density \cite{ArcherEvansDDFT,tschopp4}
\begin{align} \label{two-body exact}
\frac{1}{D_0} \, & \frac{\partial \rho^{(2)}(\vec{r}_1,\vec{r}_2,t)}{\partial t}
=
\sum_{i=1,2} \nabla_{\vec{r}_i} \cdot \Bigg(\! \nabla_{\vec{r}_i} \rho^{(2)}(\vec{r}_1,\vec{r}_2,t)
\notag\\
&\quad + \rho^{(2)}(\vec{r}_1,\vec{r}_2,t) \nabla_{\vec{r}_i} \beta \big(V_{\text{ext}}(\vec{r}_i)+\phi(r_{12})\big)
\notag\\
&\quad + \!\int d \vec{r}_3 \, \rho^{(3)}(\vec{r}_1,\vec{r}_2,\vec{r}_3,t) \nabla_{\vec{r}_i} \beta \phi(r_{i3}) \!\Bigg).
\end{align}
This equation can also be rewritten in the following alternative form
\begin{align} \label{two-body exact alternative}
\frac{1}{D_0} \, & \frac{\partial \rho^{(2)}(\vec{r}_1,\vec{r}_2,t)}{\partial t}
=
\notag\\
&\sum_{i=1,2} \nabla_{\vec{r}_i} \cdot \Bigg(\! \rho^{(2)}(\vec{r}_1,\vec{r}_2,t) \bigg(\! \nabla_{\vec{r}_i} \ln\left(\rho^{(2)}(\vec{r}_1,\vec{r}_2,t)\right)
\notag\\
&\quad +  \nabla_{\vec{r}_i} \beta \big(V_{\text{ext}}(\vec{r}_i)+\phi(r_{12})\big)
\notag\\
&\quad + \!\int d \vec{r}_3 \, \frac{\rho^{(3)}(\vec{r}_1,\vec{r}_2,\vec{r}_3,t)}{\rho^{(2)}(\vec{r}_1,\vec{r}_2,t)} \nabla_{\vec{r}_i} \beta \phi(r_{i3}) \!\bigg) \!\Bigg),
\end{align}
which remains fully equivalent to equation \eqref{two-body exact}, since there is no approximation involved at this stage.

Using the standard form \eqref{two-body exact} of the exact equation of motion and applying the adiabatic approximation solely on the three-body density, i.e. $\rho^{(3)} \!\rightarrow\! \rho^{(3)}_{\text{ad}}$, yields
\begin{align}\label{two-body adiabatic approx intermediate step}
\frac{1}{D_0} \, & \frac{\partial \rho^{(2)}(\vec{r}_1,\vec{r}_2,t)}{\partial t}
=
\sum_{i=1,2} \nabla_{\vec{r}_i} \cdot \Bigg(\! \nabla_{\vec{r}_i} \rho^{(2)}(\vec{r}_1,\vec{r}_2,t)
\notag\\
&\quad + \rho^{(2)}(\vec{r}_1,\vec{r}_2,t) \nabla_{\vec{r}_i} \beta \big(V_{\text{ext}}(\vec{r}_i)+\phi(r_{12})\big)
\notag\\
&\quad + \!\int d \vec{r}_3 \, \rho^{(3)}_{\text{ad}}(\vec{r}_1,\vec{r}_2,\vec{r}_3,t) \nabla_{\vec{r}_i} \beta \phi(r_{i3}) \!\Bigg),
\end{align}
where the adiabatic three-body density is defined as

\begin{equation}\label{rho3 ad}
\rho^{(3)}_{\text{ad}}(\vec{r}_1,\vec{r}_2,\vec{r}_3,t)
\equiv
\rho^{(3)}_{\text{eq}}(\textbf{r}_1,\textbf{r}_2,\textbf{r}_3;
[\rho(\textbf{r},t)]), 
\end{equation}
in analogy to equation \eqref{rho2 ad}.
Substitution of the second-order Yvon-Born-Green (YBG2) equation 
\cite{Hansen06,tschopp4},
\begin{align} \label{YBG 2}
\int d \vec{r}_3 \, &\rho_{\text{ad}}^{(3)}(\vec{r}_1,\vec{r}_2,\vec{r}_3, t) \nabla_{\vec{r}_i} \beta \phi(r_{i3}) =
- \nabla_{\vec{r}_i} \rho_{\text{ad}}^{(2)}(\vec{r}_1,\vec{r}_2, t)
\notag\\
& - \rho_{\text{ad}}^{(2)}(\vec{r}_1,\vec{r}_2, t) \nabla_{\vec{r}_i} 
\beta \big(V_{\text{ad}}(\vec{r}_i,t)+\phi(r_{12})\big),
\end{align}
into \eqref{two-body adiabatic approx intermediate step} yields the second equation of superadiabatic-DDFT, namely equation \eqref{two body adiabatic}. 
This is the procedure followed in reference \cite{tschopp4}.

An alternative method to implement the adiabatic closure, is to start with the (reformulated) exact 
equation of motion \eqref{two-body exact alternative} and to apply the approximation to the integral term,
\begin{align} \label{two-body alternative adiabatic approx intermediate step}
\frac{1}{D_0} \, & \frac{\partial \rho^{(2)}(\vec{r}_1,\vec{r}_2,t)}{\partial t}
=
\notag\\
&\sum_{i=1,2} \nabla_{\vec{r}_i} \cdot \Bigg(\! \rho^{(2)}(\vec{r}_1,\vec{r}_2,t) \bigg(\! \nabla_{\vec{r}_i} \ln\left(\rho^{(2)}(\vec{r}_1,\vec{r}_2,t)\right)
\notag\\
&\quad +  \nabla_{\vec{r}_i} \beta \big(V_{\text{ext}}(\vec{r}_i)+\phi(r_{12})\big)
\notag\\
&\quad + \!\int d \vec{r}_3 \, \frac{\rho^{(3)}_{\text{ad}}(\vec{r}_1,\vec{r}_2,\vec{r}_3,t)}{\rho^{(2)}_{\text{ad}}(\vec{r}_1,\vec{r}_2,t)} \nabla_{\vec{r}_i} \beta \phi(r_{i3}) \!\bigg) \!\Bigg).
\end{align}
We note that equation \eqref{two-body alternative adiabatic approx intermediate step} is no longer equivalent to the previous equation \eqref{two-body adiabatic approx intermediate step}. The essential difference between these two options is that in equation \eqref{two-body adiabatic approx intermediate step} we approximate the joint probability density, whereas in equation 
\eqref{two-body alternative adiabatic approx intermediate 
step} we approximate the conditional probability density.  
Substitution of the rewritten YBG2 equation,
\begin{align} \label{YBG 2 alternative}
&\int d \vec{r}_3 \, \frac{\rho^{(3)}_{\text{ad}}(\vec{r}_1,\vec{r}_2,\vec{r}_3,t)}{\rho^{(2)}_{\text{ad}}(\vec{r}_1,\vec{r}_2,t)} \nabla_{\vec{r}_i} \beta \phi(r_{i3}) =
\\
&
- \nabla_{\vec{r}_i} \ln\left(\rho_{\text{ad}}^{(2)}(\vec{r}_1,\vec{r}_2, t)\right)
- \nabla_{\vec{r}_i} 
\beta \big(V_{\text{ad}}(\vec{r}_i,t)+\phi(r_{12})\big),\notag
\end{align}
into expression \eqref{two-body alternative adiabatic approx intermediate step}, then yields the following alternative equation of motion for the two-body density
\begin{align} \label{two body adiabatic alternative}
&\frac{1}{D_0} \, \frac{\partial \rho^{(2)}(\vec{r}_1,\vec{r}_2,t)}{\partial t}
=
\\
&\sum_{i=1,2} \nabla_{\vec{r}_i} \cdot \Bigg(\! \rho^{(2)}(\vec{r}_1,\vec{r}_2,t) \bigg(\! \nabla_{\vec{r}_i} \ln\left(\rho^{(2)}(\vec{r}_1,\vec{r}_2,t)\right)
\notag\\
&\!+  \nabla_{\vec{r}_i} \beta \big(V_{\text{ext}}(\vec{r}_i)\!-\!V_{\text{ad}}(\vec{r}_i,t)\big)
- \nabla_{\vec{r}_i} \ln\left(\rho_{\text{ad}}^{(2)}(\vec{r}_1,\vec{r}_2, t)\right)\!
\!\bigg) \!\Bigg).\notag
\end{align}
We refer to this approximation strategy as `Russel-Gast-like', 
since equation \eqref{two body adiabatic alternative} reduces 
to the Russel-Gast equation (see reference \cite{RusselGast}) for the pair distribution function in the bulk limit, as we will demonstrate in the following subsection.

\subsection{Low shear-rate solutions for the pair distribution function in bulk}

Following the same scheme as in subsection 
\ref{super-DDFT-solutions}, the bulk limit 
of equation \eqref{two body adiabatic alternative} 
is given by
\begin{multline}
\frac{\partial g(\vec{r},t)}{\partial t} = 
-2 D_0 \nabla\cdot\Big(
-\nabla g_{\text{sup}}(\vec{r},t) +
\frac{g_{\text{sup}}(\vec{r},t)}{g_{\text{eq}}(r)}\nabla g_{\text{eq}}(r)
\Big)
\\
-\nabla\cdot\Big(
g(\vec{r},t)\,\vec{v}(\vec{r},t)
\Big).
\label{pair distribution eom alternative}
\end{multline}
This provides an alternative to equation \eqref{pair distribution eom}.
(As a side-note we mention that 
the second term on the right hand-side of equation \eqref{pair distribution eom alternative} can be rewritten using the following rearrangement
\begin{align*}
\frac{g_{\text{sup}}(\vec{r},t)}{g_{\text{eq}}(r)}\nabla g_{\text{eq}}(r) &= - g_{\text{sup}}(\vec{r},t) \nabla \Big(\! -\ln\big( g_{\text{eq}}(r) \big)  \!\Big) \\
&= - g_{\text{sup}}(\vec{r},t) \nabla \Big( \beta \phi_{\text{mf}}(r) \Big),
\end{align*}
where we have introduced the `potential of mean force', $\phi_{\text{mf}}(r)$, defined by $g_{\text{eq}}(r) \!\equiv\! e^{-\beta \phi_{\text{mf}}(r)}$.)
In the low density limit, $g_{\text{eq}}(r) \!\rightarrow\! e^{-\beta \phi(r)}$, which yields
\begin{equation}
\frac{g_{\text{sup}}(\vec{r},t)}{g_{\text{eq}}(r)}\nabla g_{\text{eq}}(r) \rightarrow - g_{\text{sup}}(\vec{r},t)\nabla\beta\phi(r),
\end{equation}
such that equations \eqref{pair distribution eom alternative} and \eqref{pair distribution eom} then become equivalent (which is not the case in general).

In the steady-state at finite densities, equation \eqref{pair distribution eom alternative} reduces to the condition
\begin{multline}
2 D_0 \nabla\cdot\Big(
\nabla g_{\text{sup}}(\vec{r},t) -
\frac{g_{\text{sup}}(\vec{r},t)}{g_{\text{eq}}(r)}\nabla g_{\text{eq}}(r)
\Big)
\\
-\nabla\cdot\Big(
g(\vec{r},t)\,\vec{v}(\vec{r},t)
\Big)=0.
\label{steady state alternative}
\end{multline}
Using the definition \eqref{split} and again assuming that 
$g_{\text{sup}}$ is linear in the 
flow-rate yields the following expression to 
leading order
\begin{multline}
2 D_0 \nabla\cdot\Big(
\nabla g_{\text{sup}}(\vec{r},t) -
\frac{g_{\text{sup}}(\vec{r},t)}{g_{\text{eq}}(r)}\nabla g_{\text{eq}}(r)
\Big)
\\
-\nabla\cdot\Big(
g_{\text{eq}}(r)\,\vec{v}(\vec{r})
\Big)=0.
\label{steady state linear alternative}
\end{multline}
Since equations \eqref{steady state linear} and 
\eqref{steady state linear alternative} are non-equivalent 
the solution of equation \eqref{steady state linear alternative} 
now requires a different ansatz than used previously. 
The appropriate choice in the present case is given by
\begin{equation}
g_{\text{sup}}(\vec{r})=
-\frac{1}{2 D_0}\left(
\frac{\vec{r}\cdot\bold{E}\cdot\vec{r}}{r^2}
\right)
g_{\text{eq}}(r) f^{\star}(r),
\label{ansatz alternative}
\end{equation}
where the $\star$ notation makes it explicit that the (still unknown) radial function, $f^{\star}(r)$, is different from the function, $f(r)$, previously encountered. 
Substitution of ansatz \eqref{ansatz alternative} into the linearized steady-state equation \eqref{steady state linear alternative} then yields the alternative radial balance equations
\begin{align}\label{radial balance alternative eq}
\frac{dg_{\text{eq}}(r)}{dr} & \, \overset{\text{2D}}{=} \,
-\frac{1}{r^2} \frac{d}{dr} \Big( r \frac{d f^{\star}(r)}{dr} g_{\text{eq}}(r) \Big) \;\,+ \frac{4}{r^3} f^{\star}(r) g_{\text{eq}}(r), \notag\\
\frac{dg_{\text{eq}}(r)}{dr} & \, \overset{\text{3D}}{=} \,
-\frac{1}{r^3} \frac{d}{dr} \Big( r^2 \frac{d f^{\star}(r)}{dr} g_{\text{eq}}(r) \Big) + \frac{6}{r^3} f^{\star}(r) g_{\text{eq}}(r),
\end{align}
to determine $f^{\star}(r)$, for the cases of two- and three-dimensions. 
We refer the reader to appendix \ref{radial balance alternative equ appendix} for additional details regarding the derivation of equations \eqref{radial balance alternative eq} and the boundary conditions required to solve them.

\subsection{Low-shear viscosity}

Within the alternative adiabatic closure the low-shear viscosity is given by 
\begin{equation}\label{viscosity alternative eq}
\eta_0 = \frac{k_BT}{4 D_0}\,\rho_b^2\int \!d\vec{r}
\frac{x^2y^2}{r^3}
\left(\frac{d}{dr}e^{-\beta\phi(r)}\right)
y_{\text{eq}}(r)
f^{\star}(r),
\end{equation}
where we have introduced the so-called `cavity distribution function',
\begin{equation}
y_{\text{eq}}(r) = g_{\text{eq}}(r)\, e^{\beta\phi(r)},
\end{equation}
familiar from liquid-state theory \cite{Hansen06}.
We emphasize that $f^{\star}(r)$ appearing in equation 
\eqref{viscosity alternative eq} is given by 
solution of the radial balance equations 
\eqref{radial balance alternative eq} and 
is distinct from the function $f(r)$ obtained from solution of 
equations \eqref{radial balance eq}. 
In two- and three-dimensions equation \eqref{viscosity alternative eq} becomes
\begin{align}\label{explicit viscosity alternative eq}
\eta_0 & \, \overset{\text{2D}}{=} \;
\frac{k_BT}{\pi D_0}\,\Phi_{\text{2D}}^2 
\int_0^{\infty}\! dr\,r^2
\left(\frac{d}{dr}e^{-\beta\phi(r)}\right)
y_{\text{eq}}(r)
f^{\star}(r), \notag\\
\eta_0 & \, \overset{\text{3D}}{=} \;
\frac{k_BT}{\pi D_0}\,\frac{12}{5}\,
\Phi_{\text{3D}}^2 
\int_0^{\infty}\! dr\,r^3
\left(\frac{d}{dr}e^{-\beta\phi(r)}\right)
y_{\text{eq}}(r)
f^{\star}(r),
\end{align}
respectively. 
For the special case of low density hard-spheres in three-dimensions, the analytic solution of equation \eqref{radial balance alternative eq} is the same as that given by 
equation \eqref{radial balance eq}, namely 
\begin{equation}
f^{\star}(r)\!=\!f(r)\!=\!1/(3r^3),
\end{equation}
such that the Russel-Gast-type closure reproduces the exact 
low density expansion of the low-shear viscosity of 
hard-spheres, given by equation \eqref{einstein eq}.

\subsection{Numerical results}\label{RG-results}

We now investigate some numerical predictions of the alternative 
closure, where we again consider the hard-disk system in two-dimensions and use the Percus-Yevick closure to obtain $g_{\text{eq}}$ for input
to the radial balance equation \eqref{radial balance alternative eq}.
\begin{figure}
\includegraphics[width=\linewidth]{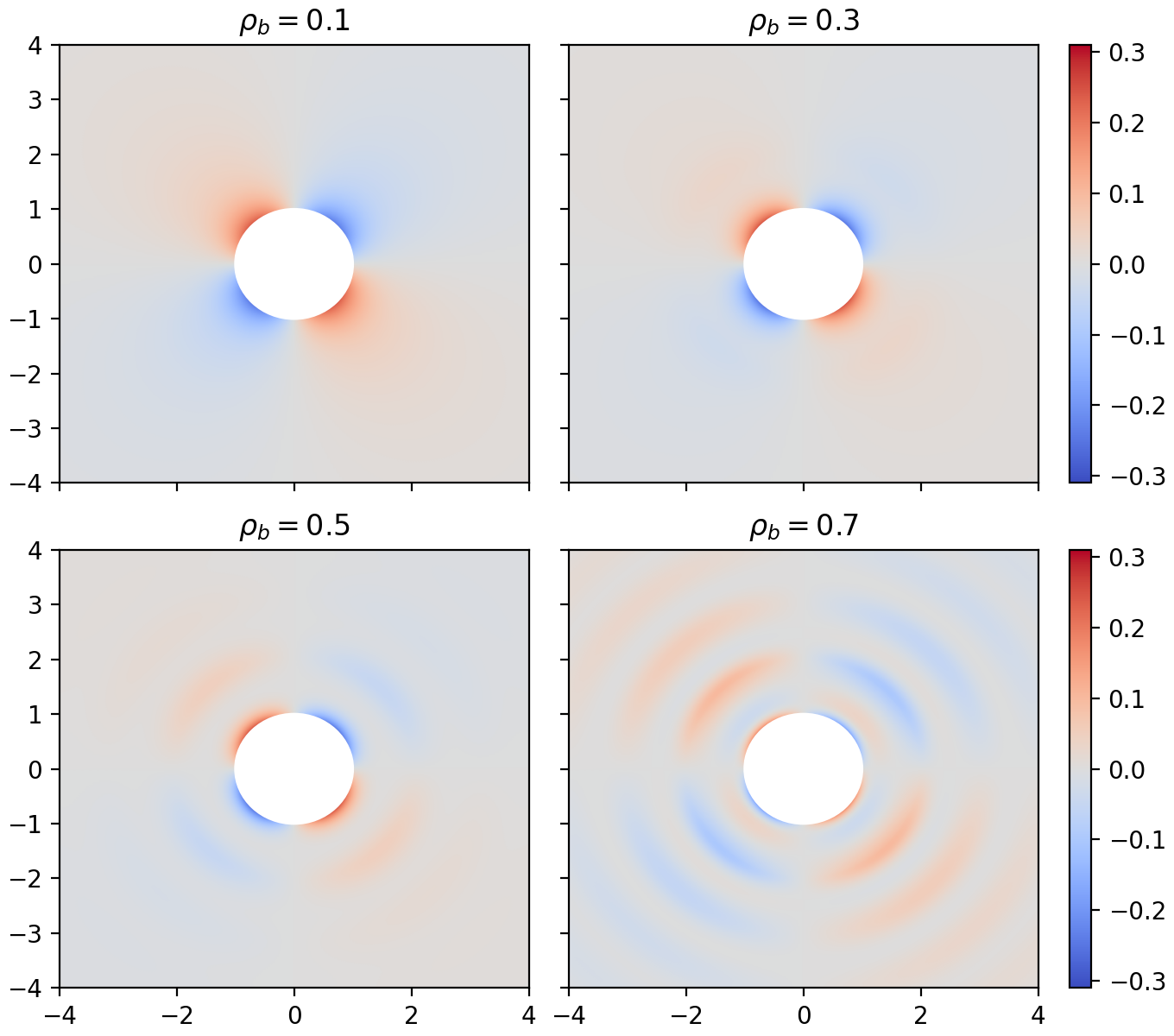}
\caption{ \textbf{Russel-Gast-type closure for \normalsize{$g_{\text{sup}}$}.}
Analogous plot to Fig.~\ref{fig Sup-DDFT}, but this time employing the alternative closure. 
For hard-disks we show
the superadiabatic contribution to the pair 
correlation function, $g_{\text{sup}}$, in units 
of $\dot{\gamma} d^2/2 D_0$, see equation 
\eqref{ansatz alternative}.
Since the results are qualitatively similar to those 
in Fig.~\ref{fig Sup-DDFT} and thus difficult to compare, we refer the reader to Fig.~\ref{fig g of r and theta} for a more detailed analysis. 
} 
\label{fig RG}
\end{figure}
In Fig.~\ref{fig RG} we show scatter plots of the quantity
\begin{equation}\label{RG viscosity}
\frac{g_{\text{sup}}(x,y)}{\dot{\gamma} d^2/2 D_0} = -\frac{xy}{r^2}f^{\star}(r) g_{\text{eq}}(r),
\end{equation}
for four different values of the input bulk density.
Note that the radial function $f^{\star}(r)$ is obtained from numerical integration of equation \eqref{radial balance alternative eq} and that $r\!=\!\sqrt{x^2+y^2}$. 
The predictions of the Russel-Gast-type closure are generally 
very similar 
to those obtained from the superadiabatic-DDFT approach of the previous section, although deviations become apparent on 
closer inspection.

The results for $g_{\text{sup}}$ at the lowest bulk density considered, $\rho_{b}\!=\!0.1$, are very close to the known low-density limit, for which 
$f^{\star}(r)\!=\!1/2r^2$, with accumulation and depletion 
at contact within the compressional and extensional quadrants, 
respectively. 
As the bulk density is increased we observe the emergence of packing oscillations. 
We recall that superadiatic-DDFT predicted that the radial function $f$ 
appearing in equation \eqref{superDDFT viscosity} has a contact 
value which remains independent of the bulk density. 
The analogous quantity within the present approximation is 
the product $f^{\star}(r)\,g_{\text{eq}}(r)$ appearing 
on the right hand-side of equation \eqref{RG viscosity}. 
We find that the contact value of this product does exhibit a nontrivial 
dependence on $\rho_b$ and, as we will see, this causes the 
low-shear viscosity generated by equation \eqref{explicit viscosity alternative eq} to deviate from that predicted by superadiabatic-DDFT.

\begin{figure}
\includegraphics[width=\linewidth]{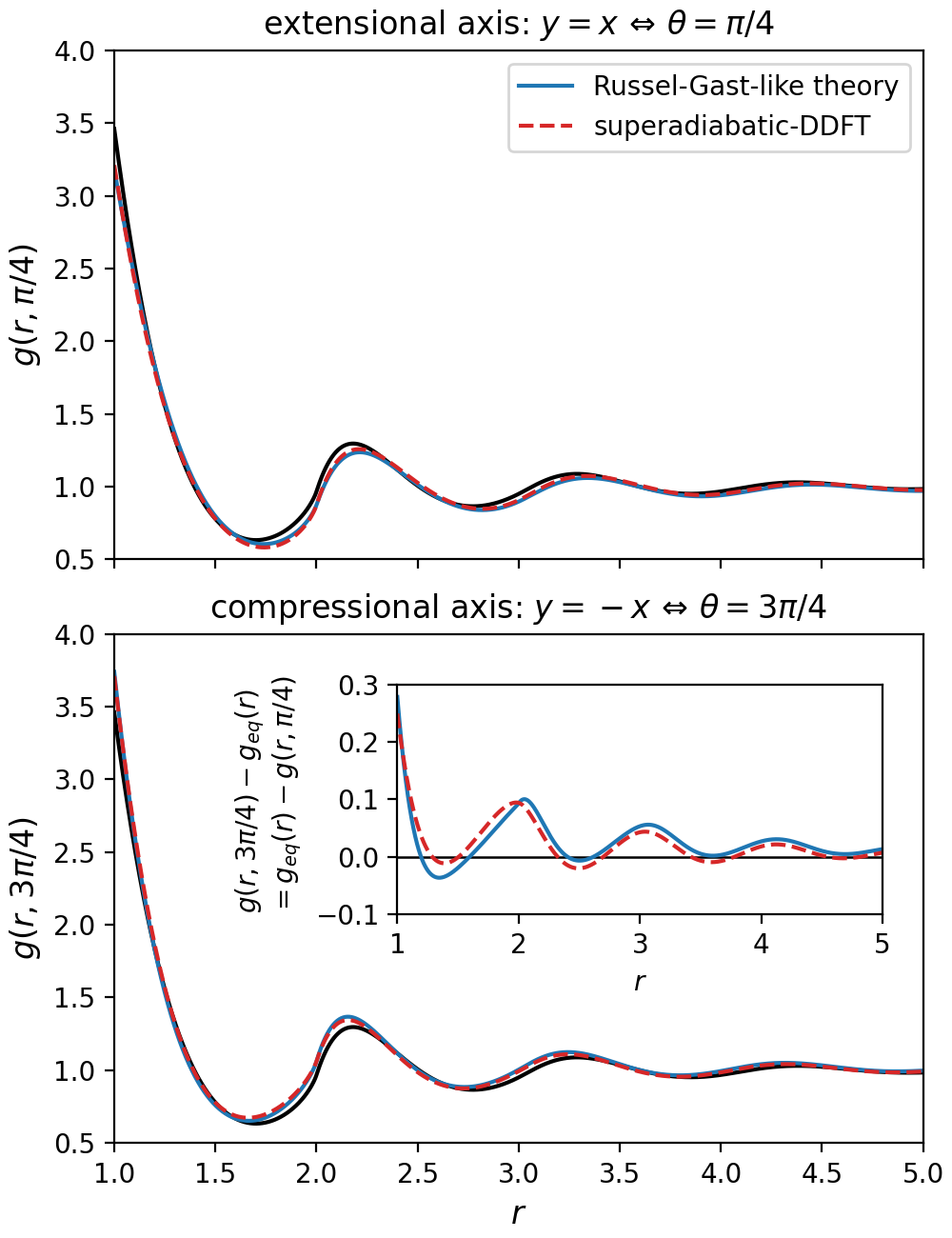}
\caption{{\bf Pair distribution function}. 
We show the full 
nonequilibrium pair distribution function of hard-disks
along the 
extensional axis, $y\!=\!x$, and the compressional 
axis, $y\!=\!-x$. 
Results are given from both the superadiabatic-DDFT 
and the Russel-Gast-type approximations, for $\rho_b\!=\!0.7$.
The additional black curve indicates the equilibrium radial distribution function for comparison. 
The dimensionless shear-rate is chosen to be 
$\dot{\gamma} d^2/2 D_0\!=\!1$. 
{\joe The inset shows the difference between the nonequilibrium  and equilibrium pair distribution 
functions to highlight the difference between the two 
approximations. Due to the symmetry of the low-shear pair distribution function for a given approximation 
the curves in the compressional and extensional 
directions are identical (up to a sign).
}}
\label{fig g of r and theta}
\end{figure}

In Fig.~\ref{fig g of r and theta} we show the 
pair distribution function on both the extensional axis, along which the particles get pulled apart by the shear, and on the compressional axis, along which the particles get pressed together. 
{\joe In order to show more clearly the difference between the superadiabatic-DDFT and Russel-Gast-type 
approximations we show in the inset the difference between the nonequilibrium and equilibrium pair distribution functions.}
On the extensional axis both approximation schemes predict 
a reduction in the height of all maxima, including 
the first peak at particle contact. 
The reduction in amplitude of the peaks is slightly more pronounced within the Russel-Gast-type approximation. 
In both cases the radial position of the second and higher-order peaks shifts to larger values of $r$, consistent with the fact that particles are being pulled away from each other 
by the shear flow.  
In contrast, on the compressional axis the height of all maxima increase and the radial locations of the second and higher-order peaks are shifted to smaller values of $r$, since the particles are being pushed closer together by the shear flow. 
The two theories again make similar predictions, although 
the Russel-Gast-type approximation yields a slightly larger 
increase in peak height than the superadiabatic-DDFT.
The general similarity of the predictions from the two approximation schemes for the nonequilibrium microstructure 
is reassuring, as it appears that the results are robust 
with respect to the details of how the adiabatic approximation 
is implemented. 

\begin{figure}
\includegraphics[width=\linewidth]{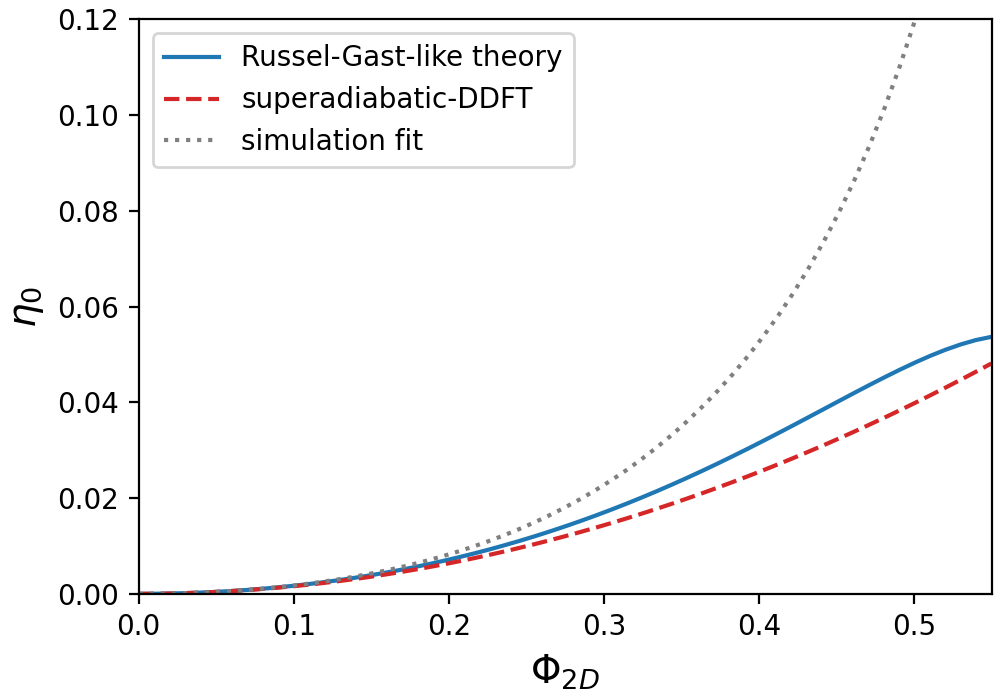}
\caption{\textbf{Low-shear viscosity}. 
Analogous plot to Fig.~\ref{fig Sup-DDFT viscosity}, but now 
including the prediction of the Russel-Gast-type approximation.
This additional result predicts slightly larger values of $\eta_0$  than the superadiabatic-DDFT, but exhibits an unphysical curvature as the area fraction is increased. 
}
\label{fig viscosity RG and Sup-DDFT}
\end{figure}

From the nonequilibrium pair correlation function we can calculate the zero-shear viscosity using equation 
\eqref{explicit viscosity alternative eq}, for which  
results are shown in Fig.~\ref{fig viscosity RG and Sup-DDFT}. 
We find that the Russel-Gast-type approximation scheme 
generates a low-shear viscosity curve with values slightly larger than that 
predicted by the superadiabatic-DDFT. 
We can attribute this effect to the difference in the contact 
value of the pair distribution function between the two approximations, see Fig.~\ref{fig g of r and theta}. 
Within the Russel-Gast-type approach the difference between the contact value on the extensional axis and the contact value on the  compressional axis is larger than 
for superadiabatic-DDFT. 
This larger extensional/compressional asymmetry has the consequence that for any given area fraction $\eta_0$ 
from the Russel-Gast-type approach is larger than for the superadiabatic-DDFT.  
Although the Russel-Gast-type closure produces values 
of the low-shear viscosity closer to the simulation 
results, it has the undesirable feature of changing curvature as the area fraction increases beyond about $0.45$. This questions the physicality of the alternative approximation and its rescaling potential for higher densities \cite{BradyScaling}.

\section{Discussion}\label{discussion}

In this paper we have investigated the predictions of superadiabatic-DDFT for the nonequilibrium pair correlation
function and low-shear viscosity of bulk systems subject to 
the homogeneous shear flow defined by equation 
\eqref{shear definition}. 
This provides a conceptual link between density 
functional-based approaches, which are focused on the inhomogeneous one-body density \cite{Marconi98,ArcherEvansDDFT,GoddardHydro2016}, 
and microscopic theories of homogeneous bulk rheology, 
of the type pioneered by Brady, Russel, Wagner and others 
\cite{Russel_1989,WagnerBook,BraderReview}. 
A clear connection between these two substantial bodies of literature, which have coexisted for decades with little to 
no interaction, is established by our Russel-Gast-type closure of the equation of motion for the two-body density. 
As detailed in section \ref{section RG}, this new scheme generates a fully inhomogeneous DDFT, which reduces to the known Russel-Gast theory for bulk systems subject to 
homogeneous shear \cite{RusselGast}. 

The application of shear distorts the bulk pair 
distribution function and, for systems interacting via 
a pairwise additive potential, enables the interaction 
part of the stress tensor to be calculated. 
For the low shear-rates considered here the only relevant transport coefficient is the viscosity, since the lowest-order 
shear-induced changes to the diagonal stress tensor elements 
are quadratic in $\dot{\gamma}$.
For larger shear-rates we would need to consider 
the extension 
of our solution ansatz \eqref{ansatz} or \eqref{ansatz alternative}, respectively, to 
higher-order in $\dot{\gamma}$, which would then generate 
nonvanishing normal stress differences and shear-induced 
modifications to the system pressure 
\cite{Bergenholtz,BraderScacchi}. 
{\joe It would be interesting to investigate the 
extent to which numerical solutions of 
the superadiabatic-DDFT can account for nonlinear rheological 
and microstructural changes as the shear rate is increased. 
For example, it is known from both simulation 
\cite{BradyTripletSimulations} and experiment \cite{Vermant_2005} that the nonequilibrium bulk 
three-body distribution function deviates significantly 
from its equilibrium form, which could be difficult to 
capture using an adiabatic closure on the two-body level.}

Focusing on the low shear-rate regime enabled a largely analytic 
investigation of the nonequilibrium pair distribution function, 
with only the solution of the radial balance equation and 
evaluation of the low-shear viscosity demanding (relatively 
simple) numerical integration.  
At larger values of the shear-rate more sophisticated techniques must be employed, since boundary-layer 
formation prevents the application of straightforward perturbation theory \cite{Dhont_1989,Bergenholtz,Bergenholtz_review,
zaccone_intermediate,riva}. 
An interesting extension of our study to obtain results 
for higher shear-rates could be to employ either 
multipole methods \cite{Szamel_1993}, specialized numerical schemes \cite{Bergenholtz,Nazockdast_Morris_2012} or intermediate asymptotics \cite{banetta,riva}.
Although the theories presented here are valid for an arbitrary 
pair interaction potential in two- or three-dimensions, we chose to perform numerical calculations for the special case 
of hard-disks, as this presents 
the simplest continuum model for the study of 
shear, while still retaining much of the essential 
phenomenology of three-dimensional systems. 
In references \cite{banetta,riva} it is shown how 
the pair distribution function under shear can be calculated 
analytically for more general interaction potentials.

The two routes presented in this work are built on two variations of the same approximation. The numerical output of those theories is broadly similar but, as clearly pictured by the final viscosity plot, Fig.~\ref{fig viscosity RG and Sup-DDFT}, are different in more demanding situations at higher bulk density. 
One can thus ask which scheme is more likely to succeed in further investigations. 
The unphysical shape of the viscosity curve obtained with the Russel-Gast-like closure suggests that this variation of the approximation may be less robust than that of superadiabatic-DDFT and could prove more difficult to improve and refine. 

A natural first step towards better predictions for the viscosity at higher bulk densities would be to apply Brady's semi-empirical rescaling method \cite{BradyScaling}.
For the present Brownian dynamics, without hydrodynamic interactions, this consists of multiplying the 
low density limiting solution for the interaction contribution 
to $\eta_0$ (given by equation \eqref{superDDFT viscosity} 
for hard-disks) by the contact value of the equilibrium radial distribution function. 
The resulting low-shear viscosity has been shown to agree 
well with simulation 
data for hard-spheres \cite{LionbergerRusselReview} and we find that this is also the case for hard-disks. 
Since the superadiabatic-DDFT predicts that 
the low-density limiting result \eqref{superDDFT viscosity} holds for all values of $\rho_b$, a rescaled version of 
superadiatic-DDFT would exactly reproduce Brady's theory 
for $\eta_0$.
In contrast, a rescaled Russel-Gast-type theory would 
continue to exhibit an unphysical curvature, as the contact value of $g_{\text{eq}}$ is a monotonically increasing function of bulk density.  
Investigating how such a rescaling factor could emerge 
from a systematic, first-principles extension of the superadiabatic-DDFT closure 
scheme will be an interesting topic for future study. 

{\joe
While the above considerations may provide a path to structural improvements of the 
superadiabatic-DDFT equations (i.e.~following from a new 
two-body closure) 
there remains a great deal to be learned from the current 
version of the theory.  
In this paper we have considered only the 
{\it direct application} of superadiabatic-DDFT to the bulk system, meaning that we work from the outset with a constant 
one-body density and focus on the properties and predictions 
of the resulting two-body equation of motion.
Although this sheds light on the internal structure 
and physical content of the theory, we should keep in mind 
that the superadiabatic-DDFT is still a density functional 
theory and, as such, aims primarily to predict the 
dynamics of the one-body density.   
If we are interested solely in bulk systems under shear, 
as in the present work, 
then the 
full power of superadiabatic-DDFT for the one-body dynamics 
can be harnessed by employing a more sophisticated implementation scheme: the test-particle method.  
Fixing a test-particle at the coordinate origin (i.e.~setting 
$V_{\text{ext}}(\vec{r})\!=\!\phi(r)$) induces a 
spatially varying steady-state one-body density, 
$\rho(\vec{r})$,  
as particles around the test-particle accumulate in 
the compressional quadrants and are depleted from the extensional quadrants \cite{Reinhardt_2013}. This inhomogeneous one-body density can be related 
to the {\it bulk} pair distribution function according to 
$g(\vec{r})\!=\!\rho(\vec{r})/\rho_b$, from which 
the viscosity can be calculated using equation \eqref{stress eq}. 
We suspect that a test-particle implementation 
of superadiabatic-DDFT will yield a low-shear viscosity which much improves on the simple quadratic expression 
\eqref{superDDFT viscosity}, obtained by direct 
application of the two-body equation of motion to bulk. 
The proposed test-particle calculation would require 
explicit treatment of the inhomogeneous two-body 
density in the presence of a 
test-particle and would thus capture, 
to some extent, the shear-induced distortion of the bulk 
three-body correlations. 
Investigations in this direction are underway. 
}
\\



Finally, we mention the issue of hydrodynamic interactions.  
Although the majority of standard DDFT studies do not 
consider solvent hydrodynamics, this aspect has been incorporated 
into the formalism 
\cite{rex,rex2,RauscherHydro,Goddard_2013,GoddardArcher}. 
It would thus be interesting to investigate whether a similar extension would be feasable for superadiabatic-DDFT. 
Many of the existing theories of homogeneous bulk rheology 
mentioned in the introduction, including the bulk Russel-Gast theory, were formulated to include hydrodynamic interactions to some level of approximation. These works, all 
of which are focused on the two-body correlations, could 
well provide a source of inspiration for future developments.

\appendix

\section{Rheological quantities}\label{rheology appendix}

In a system undergoing translationally invariant, homogeneous flow, the velocity field can be conveniently expressed 
in the following form
\begin{equation}\label{velocity gradient}
\vec{v}(\vec{r}) = 
\boldsymbol{\kappa}\cdot\vec{r}, 
\end{equation}
where $\boldsymbol{\kappa}$ is the spatially constant 
velocity gradient tensor. 
As an example, choosing flow in the $x$-direction, 
with shear-gradient in the $y$-direction enables 
$\boldsymbol{\kappa}$ to be represented in matrix form 
as 
\begin{equation}
\boldsymbol{\kappa} = 
\begin{pmatrix}
0 & \dot{\gamma} & 0 \\
0 & 0 & 0 \\
0 & 0 & 0
\end{pmatrix},
\end{equation}
for a three-dimensional system with shear-rate $\dot{\gamma}$.

The right hand-side of \eqref{velocity gradient} can be decomposed into a sum of two terms,
\begin{equation}\label{decomposition}
\vec{v}(\vec{r}) = 
\E\cdot\vec{r} + \Om\cdot\vec{r}, 
\end{equation}
where 
\begin{equation}\label{E and Om}
\E=\frac{1}{2}\left( \kap + \kap^{\text{T}} \right),
\quad
\Om=\frac{1}{2}\left( \kap - \kap^{\text{T}} \right), 
\end{equation}
are the (symmetric) rate-of-strain tensor and 
the (antisymmetric) rate-of-rotation 
tensor, respectively.
The rate-of-strain tensor describes `pure straining motion' leading to relative motion between any two particles, whereas the rate-of-rotation tensor 
yields pure rotational motion, which does not affect their relative separation \cite{dhont}. 
For the aforementioned case of shear flow we 
obtain
\begin{equation}\label{E matrix}
\boldsymbol{\E} = 
\frac{1}{2}
\begin{pmatrix}
0 & \dot{\gamma} & 0 \\
\dot{\gamma} & 0 & 0 \\
0 & 0 & 0
\end{pmatrix}, 
\quad
\boldsymbol{\Om} = 
\frac{1}{2}
\begin{pmatrix}
\;\;0 & \dot{\gamma} & 0 \\
\!\!\!-\dot{\gamma} & 0 & 0 \\
\;\;0 & 0 & 0
\end{pmatrix}.
\end{equation}
The anisotropy of the nonequilibrium pair distribution 
function in a system subject to a slow translationally invariant flow is determined solely by its straining 
motion
and does not involve its rotational component 
\cite{Russel_1989,Bergenholtz}. 
This motivates our choice of ansatz for $g_{\text{sup}}$ in 
equations \eqref{ansatz} and 
\eqref{ansatz alternative}.

\section{The radial balance equation}\label{radial balance equ appendix}

Substitution of the ansatz \eqref{ansatz} into the linearized 
steady-state equation \eqref{steady state linear} yields the radial balance equations \eqref{radial balance eq}.
Since this procedure is not straightforward we outline here the main steps of the calculation. 
Although we are primarily interested in the case of homogeneous shear, we formulate the problem using the rate-of-flow tensor $\E$, which can also describe other flow types. 
This not only extends the generality of our results, but also presents the technical advantage that certain vector/tensor identities can be employed 
to make the calculation as clean as possible. Unless otherwise stated, all expressions given in this appendix are valid in arbitrary dimensionality.

We start by collecting a few useful identities. 
For a general, spatially varying, second-rank tensor $\M$ the following holds 
\begin{align}\label{identity_first}
\nabla\big(
\left(\hat{\vec{r}}\cdot\M\cdot\hat{\vec{r}}\right)\,a(r)
\big)
&=\frac{2}{r^{2}}(\vec{r}\cdot\M)\cdot(\unittensor-\hat{\vec{r}}\hat{\vec{r}})\,a(r)
\notag\\
&+\frac{1}{r}\frac{da(r)}{dr}(\vec{r}\cdot\M)\cdot\hat{\vec{r}}\hat{\vec{r}}
,
\end{align}
where $\hat{\vec{r}}\!=\!\vec{r}/r$ is a unit vector, 
$\hat{\vec{r}}\hat{\vec{r}}$ is a dyadic product and where 
$a(r)$ is a scalar function of $r\!=\!|\vec{r}|$.
For an arbitrary, spatially dependent vector field, $\vec{u}$, we have 
\begin{equation}
\label{identity_second}
\nabla\cdot\left(\textbf{u}\cdot\bold{M}\right)=\textbf{u}\cdot\left(\nabla\cdot\bold{M}\right)
+\bold{M}:\nabla\textbf{u},
\end{equation}
where the double dot notation in the second term 
indicates a full contraction, i.e.~a scalar product 
followed by a trace operation.
The divergence of the product $a(r)\M$ is given by
\begin{equation}
\label{identity_third}
\begin{split}
\nabla\cdot\left(a(r)\bold{M}\right)&=\nabla a(r)\cdot\bold{M}+a(r)\,\nabla\cdot\bold{M}
\\
&=\left(\frac{da(r)}{dr}\right)\hat{\textbf{r}}\cdot\bold{M}+a(r)\,\nabla\cdot\bold{M}.
\end{split}
\end{equation}
This identity generates the following 
special cases in two-dimensions
\begin{align}
\label{identity_fourth}
&\nabla\cdot\big(\,a(r)\left(\unittensor-\hat{\textbf{r}}\hat{\textbf{r}}\right)\big)
\overset{\text{2D}}{=}
-\frac{a(r)}{r}\,\hat{\vec{r}}, 
\\
&\nabla\cdot\big(\,a(r)\,\hat{\textbf{r}}\hat{\textbf{r}}
\,\big)
\overset{\text{2D}}{=}\frac{1}{r}\frac{d}{dr}
\Big(
r a(r)
\Big)
\hat{\vec{r}},
\label{identity_fourthB}
\end{align}
and three-dimensions
\begin{align}
\label{identity_fourth 3d}
&\nabla\cdot\big(\,a(r)\left(\unittensor-\hat{\textbf{r}}\hat{\textbf{r}}\right)\big)
\overset{\text{3D}}{=}
-\frac{2a(r)}{r}\,\hat{\vec{r}}, 
\\
&\nabla\cdot\big(\,a(r)\,\hat{\textbf{r}}\hat{\textbf{r}}
\,\big)
\overset{\text{3D}}{=}\frac{1}{r^2}\frac{d}{dr}
\Big(
r^2 a(r)
\Big)
\hat{\vec{r}},
\label{identity_fourthB 3d}
\end{align}
respectively. 
In the special case that $\M$ is symmetric it can be shown 
that  
\begin{equation}
\big(\nabla\left(\M\cdot\hat{\vec{r}}\right)\big)\cdot\hat{\vec{r}}=0.
\label{identity_eighth}
\end{equation}
Finally, the divergence of the scalar product 
$\M\cdot\hat{\vec{r}}$ is given by the aesthetically appealing identity
\begin{equation}
\nabla\cdot\left(\M\cdot\hat{\vec{r}}\right)=\frac{{\rm Tr}\left(\M\right)}{r} 
- \frac{\hat{\vec{r}}\cdot\M\cdot\hat{\vec{r}}}{r}.
\label{identity_seventh}
\end{equation}
We will now employ these identities to calculate the 
pair distribution function at low flow-rate.

Substitution of the ansatz \eqref{ansatz} into the linearized steady-state equation \eqref{steady state linear} yields 
%
%
%
%
\begin{align}\label{linear_smol}
\nabla\cdot
\Big(g_{\text{eq}}(r)\vec{v}(\vec{r})\Big) 
= 
-\nabla\cdot\Big( e^{-\beta\phi(r)}\,
\nabla\big(
\left(\hat{\vec{r}}\cdot\E\cdot
\hat{\vec{r}}\right)f(r)
\big)
\Big),
\end{align}
in which the only unknown quantity is the function $f(r)$. 
Assuming incompressible flow, $\nabla\cdot \vec{v}\!=\!0$,
using identity \eqref{identity_first} and rewriting the projected velocity as 
\begin{equation}\label{projected velocity}
\vec{v}(\vec{r})\cdot\hat{\vec{r}}
=
r\left(
\hat{\vec{r}}\cdot\E\cdot\hat{\vec{r}}
\right),
\end{equation}
enables us to re-express equation \eqref{linear_smol} as follows
\begin{align}\label{linear sub eq}
&r\frac{dg_{\text{eq}}(r)}{dr}\,
\big(\hat{\vec{r}}\cdot\E\cdot\hat{\vec{r}}\big)
=
\\
&-\nabla\cdot
\bigg(
\big(\hat{\vec{r}}\cdot\E\big)
\cdot
\bigg(
(\unittensor-\hat{\vec{r}}\hat{\vec{r}}) \frac{2f(r)}{r}
+\hat{\vec{r}}\hat{\vec{r}}\frac{df(r)}{dr}
\bigg)
e^{-\beta\phi(r)}
\bigg).
\notag
\end{align}
The advantage of this representation is the appearance 
of the dyadic tensors, 
$\hat{\vec{r}}\hat{\vec{r}}$ and 
$(\unittensor-\hat{\vec{r}}\hat{\vec{r}})$, which project either along or perpendicular to the relative position vector $\vec{r}$.

On the right hand-side of \eqref{linear sub eq} we have to calculate the divergence of the scalar product between the 
vector $\hat{\vec{r}}\cdot\E$ and a second-rank tensor. 
We can thus exploit relation (\ref{identity_second}) 
to obtain
\begin{align}\label{simplified}
r\frac{dg_{\text{eq}}(r)}{dr}\,
\big(\hat{\vec{r}}\cdot\E\cdot\hat{\vec{r}}\big)
&=
\notag\\
&\hspace*{-2.8cm}
-\left(\hat{\vec{r}}\cdot\E\right)\cdot\bigg(\nabla\cdot\bigg(
\!\bigg(
(\unittensor-\hat{\vec{r}}\hat{\vec{r}}) \frac{2f(r)}{r}
+\hat{\vec{r}}\hat{\vec{r}}\frac{df(r)}{dr}
\bigg)
e^{-\beta\phi(r)}
\bigg)\!\bigg)
\notag\\
&\hspace*{-2.8cm}
-\left(
\!\bigg(
(\unittensor-\hat{\vec{r}}\hat{\vec{r}}) \frac{2f(r)}{r}
+\hat{\vec{r}}\hat{\vec{r}}\frac{df(r)}{dr}
\bigg)
e^{-\beta\phi(r)}
\right):\nabla\left(\E\cdot\hat{\vec{r}}\right).
\end{align}
We will consider separately the two terms appearing
on the right hand-side of \eqref{simplified}, which we henceforth refer to as $(i)$ and $(ii)$.

To simplify $(i)$ in two-dimensions we use \eqref{identity_third}, 
\eqref{identity_fourth} and \eqref{identity_fourthB}. 
This yields
\begin{equation}\label{subcalc1}
-\left(\hat{\vec{r}}\cdot\E\cdot\hat{\vec{r}}\right)
\!
\left(
\frac{1}{r}\frac{d}{dr}\left(r
\frac{df(r)}{dr}e^{-\beta\phi(r)}\!\right) 
- \frac{2f(r)e^{-\beta\phi(r)}}{r^{2}}\!
\right).
\end{equation}
The analogous result in three-dimensions is given by
\begin{equation}\label{subcalc1 3d}
-\left(\hat{\vec{r}}\cdot\E\cdot\hat{\vec{r}}\right)
\!
\left(
\frac{1}{r^2}\frac{d}{dr}\left(r^2
\frac{df(r)}{dr}e^{-\beta\phi(r)}\!\right) 
- \frac{4f(r)e^{-\beta\phi(r)}}{r^{2}}\!
\right),
\end{equation}
where we have used equations 
\eqref{identity_third}, \eqref{identity_fourth 3d} and \eqref{identity_fourthB 3d}.

The simplification of term $(ii)$ does not depend on the 
dimensionality of the system. We first employ equation 
\eqref{identity_eighth} to re-express the factor 
$\nabla(\E\cdot\hat{\vec{r}})$ as a divergence 
and then use identity \eqref{identity_seventh}
to obtain 
\begin{align}\label{subcalc2}
&-\frac{2f(r)e^{-\beta\phi(r)}}{r}\nabla\cdot\left(\bold{E}\cdot\hat{\textbf{r}}\right)
\notag\\
&\qquad=
-\frac{2f(r)e^{-\beta\phi(r)}}{r}\left(\frac{{\rm Tr}\left(\bold{E}
\right)}{r}
-\frac{\hat{\textbf{r}}\cdot\bold{E}\cdot\hat{\textbf{r}}}{r}\right)
\notag\\
&\qquad=
\frac{2f(r)e^{-\beta\phi(r)}}{r^{2}}\left(\hat{\textbf{r}}\cdot\bold{E}\cdot\hat{\textbf{r}}\right),
\end{align}
since ${\rm Tr}\left(\E\right)=0$ for incompressible flow. 

Finally, substituting \eqref{subcalc1} and \eqref{subcalc2} into \eqref{simplified} yields the desired two-dimensional 
result 
\begin{align}\label{radial appendix final 2D}
\left(\hat{\vec{r}}\cdot\E\cdot\hat{\vec{r}}\right)
&\left(\frac{1}{r}\frac{d}{dr}\left(r
\frac{df(r)}{dr}e^{-\beta\phi(r)}\right)
- \frac{4}{r^{2}}f(r)e^{-\beta\phi(r)}\right)
\notag\\
&\overset{\text{2D}}{=}-\left(\hat{\textbf{r}}\cdot\bold{E}\cdot\hat{\textbf{r}}
\right)\,r\frac{dg_{\text{eq}}(r)}{dr}, 
\end{align}
while substitution of \eqref{subcalc1 3d} and 
\eqref{subcalc2} into \eqref{simplified} gives the result in three-dimensions 
\begin{align}\label{radial appendix final 3D}
\left(\hat{\vec{r}}\cdot\E\cdot\hat{\vec{r}}\right)
&\left(\frac{1}{r^2}\frac{d}{dr}\left(r^2
\frac{df(r)}{dr}e^{-\beta\phi(r)}\right)
- \frac{6}{r^{2}}f(r)e^{-\beta\phi(r)}\right)
\notag\\
&\overset{\text{3D}}{=}-\left(\hat{\textbf{r}}\cdot\bold{E}\cdot\hat{\textbf{r}}
\right)\,r\frac{dg_{\text{eq}}(r)}{dr}.
\end{align}
Since these expressions remain valid for all choices of the traceless tensor $\E$, the coefficients of the quadratic 
form $(\hat{\vec{r}}\cdot\E\cdot\hat{\vec{r}})$ must be equal.
We thus obtain the radial balance equations 
\eqref{radial balance eq} stated in the main text. 
We note that, in addition to the explicit appearance of 
the pair interaction potential in 
\eqref{radial appendix final 2D}, respectively \eqref{radial appendix final 3D}, 
both the pair potential and the bulk density 
enter these equations implicitly via the equilibrium radial distribution 
function.

\section{Boundary conditions}\label{boundary conditions}

The radial balance equations \eqref{radial balance eq} are 
second-order in spatial derivative and their solution thus requires the specification of two boundary conditions.  
From the linearized steady-state equation 
\eqref{steady state linear} we can identify 
the following expression for the pair-current at low 
shear-rates 
\begin{equation}
\vec{j}(\vec{r}) \!=\! -2 D_0 \Big(
\nabla g_{\text{sup}}(\vec{r}) + 
g_{\text{sup}}(\vec{r})\nabla\beta\phi(r)
\Big)
\!+\!
g_{\text{eq}}(r)\,\vec{v}(\vec{r})
.
\end{equation}
Pairs of particles separated by a large distance are not spatially correlated, 
which implies that the radial component of the 
pair-current should tend to zero. 
Moreover, if we assume that the interparticle interaction potential has a strongly repulsive 
core, then we can impose that 
the radial component of the pair-current will go to zero 
also at small separations. 
We thus have the boundary condition
\begin{equation}\label{bc1 general}
\vec{j}(\vec{r})\cdot\hat{\vec{r}}=0 
\end{equation}
for both $r\!\rightarrow\!\infty$ and $r\!\rightarrow\!0$. 

For the case of hard-disks in two-dimensions or 
hard-spheres in three-dimensions, the second boundary condition takes 
a special form, since the radial pair-current must 
vanish when two particles touch in order to prevent 
unphysical overlap. 
The small separation boundary condition then becomes 
\begin{equation}\label{bc1 hs}
\vec{j}(\vec{r})\cdot\hat{\vec{r}}
\underset{r\rightarrow 1}{=}0 ,
\end{equation}
where we have set the particle diameter equal 
to unity. Using steps directly analogous to those leading from 
\eqref{linear_smol} to \eqref{linear sub eq} enables 
the pair-current of hard-spheres to be expressed in 
the following form
\begin{equation}\label{hs current}
\vec{j}(\vec{r})\!=\!
\big(\hat{\vec{r}}\cdot\E\big)
\cdot
\bigg(\!
(\unittensor-\hat{\vec{r}}\hat{\vec{r}}) \frac{2f(r)}{r}
+\hat{\vec{r}}\hat{\vec{r}}\frac{df(r)}{dr}
\bigg)
+\, g_{\text{eq}}(r)\vec{v}(\vec{r}),
\end{equation}
for $r\!>\!1$. 
(Note that $\phi(r\!>\!1)\!=\!0$ for 
the hard-sphere potential.)
Taking the scalar product of equation \eqref{hs current} with 
the radial unit vector yields
\begin{equation}
\vec{j}(\vec{r})\cdot\hat{\vec{r}}
=
(\hat{\vec{r}}\cdot
\E
\cdot\hat{\vec{r}})\left(
rg_{\text{eq}}(r) + \frac{df(r)}{dr}
\right),
\end{equation}
where we have used \eqref{projected velocity}. 
The hard-sphere `zero-flux' condition at particle contact 
thus becomes
\begin{equation}\label{hsboundary}
\frac{df(r)}{dr}\bigg|_{r=1} = -g_{\text{eq}}(1).
\end{equation}
The radial balance equations \eqref{radial balance eq}, combined with \eqref{hsboundary} and 
$f(r\!\rightarrow\!\infty)\!=\!0$ fully determine the function $f(r)$, given that the input equilibrium radial distribution function is known.

\section{Alternative forms of the radial balance equation 
and boundary conditions}\label{radial balance alternative equ appendix}

In the preceeding two appendices we provided all details 
required for the derivation and solution of the radial 
balance equations of superadiabatic-DDFT. 
Since the derivation of both the alternative 
radial balance equations \eqref{radial balance alternative eq} 
and their boundary conditions are very similar, we give here 
only the main equations to highlight the differences between the two approximation schemes.

Substitution of the ansatz \eqref{ansatz alternative} into the linearized steady-state equation \eqref{steady state linear alternative} yields 
\begin{align}\label{RG1 appendix}
\nabla\cdot
\Big(g_{\text{eq}}(r)\vec{v}(\vec{r})\Big) 
= 
-\nabla\cdot\Big( g_{\text{eq}}(r)\,
\nabla\big(
\left(\hat{\vec{r}}\cdot\E\cdot
\hat{\vec{r}}\right)f^{\star}(r)
\big)
\Big), 
\end{align}
from which we can determine the function $f^{\star}(r)$. 
Assuming incompressible flow, $\nabla\cdot \vec{v}\!=\!0$ 
and using the identities \eqref{identity_first} and 
\eqref{projected velocity} we can re-express equation \eqref{RG1 appendix} in the following form
\begin{align}\label{RG2 appendix}
&r\frac{dg_{\text{eq}}(r)}{dr}\,
\big(\hat{\vec{r}}\cdot\E\cdot\hat{\vec{r}}\big)
=
\\
&-\nabla\cdot
\bigg(
\big(\hat{\vec{r}}\cdot\E\big)
\cdot
\bigg(
(\unittensor-\hat{\vec{r}}\hat{\vec{r}}) 
\frac{2f^{\star}(r)}{r}
+\hat{\vec{r}}\hat{\vec{r}}
\frac{df^{\star}(r)}{dr}
\bigg)
g_{\text{eq}}(r)
\bigg).
\notag
\end{align}
Steps analogous to those leading from 
equation \eqref{linear sub eq} to 
\eqref{radial appendix final 3D} then generate the alternative form of the radial balance equations 
\eqref{radial balance alternative eq}. 

From the linearized steady-state equation 
\eqref{steady state linear alternative} we identify 
the pair-current at low shear-rates 
\begin{multline}\label{current linear alternative}
\vec{j}(\vec{r},t) = 
-2 D_0 \Big(
\nabla g_{\text{sup}}(\vec{r},t) -
\frac{g_{\text{sup}}(\vec{r},t)}{g_{\text{eq}}(r)}\nabla g_{\text{eq}}(r)
\Big)
\\
+
g_{\text{eq}}(r)\,\vec{v}(\vec{r}).
\end{multline}
A calculation analogous to that leading from 
\eqref{linear_smol} to \eqref{linear sub eq} 
shows that equation \eqref{current linear alternative} 
can be re-expressed as
\begin{multline}\label{current alternative}
\vec{j}(\vec{r})\!=\!
\big(\hat{\vec{r}}\cdot\E\big)
\!\cdot\!
\bigg(\!
(\unittensor-\hat{\vec{r}}\hat{\vec{r}}) 
\frac{2f^{\star}(r)}{r}
+\hat{\vec{r}}\hat{\vec{r}}\frac{df^{\star}(r)}{dr}
\bigg)g_{\text{eq}}(r)
\\
+\, g_{\text{eq}}(r)\vec{v}(\vec{r}).
\end{multline}
For a general interaction potential the boundary condition 
\eqref{bc1 general} holds for both $r\!\rightarrow\!\infty$ 
and $r\!\rightarrow\!0$. 

For the special case of hard-spheres the two-body current must be equal to zero when a pair of particles come into contact. 
Applying the boundary condition \eqref{bc1 general} at 
$r\!=\!1$ yields
\begin{equation}\label{hsboundary alternative}
\frac{df^{\star}(r)}{dr}\bigg|_{r=1} = -1,
\end{equation}
which is different from equation \eqref{hsboundary}
The boundary condition at particle contact, equation \eqref{hsboundary alternative}, together with 
$f^{\star}(r\!\rightarrow\!\infty)\!=\!0$, then fully determines 
the solutions of the radial balance equations \eqref{radial balance alternative eq}.

\bibliography{paper6}




\end{document}